\documentclass[aps,prc,twocolumn,groupedaddress,showpacs,10pt]{revtex4-1}

\usepackage{graphicx}

\begin{document}


\title{GT strengths and electron-capture rates for $pf$-shell nuclei of relevance for late stellar evolution.}

\author{A.L. Cole}
\email[]{acole@kzoo.edu}
\author{T.S. Anderson}
\affiliation{Physics Department, Kalamazoo College, Kalamazoo, MI 49006, USA}
\author{R.G.T. Zegers}\email[]{zegers@nscl.msu.edu}
\author{Sam M. Austin}
\author{B.A. Brown}
\author{L. Valdez}
\affiliation{National Superconducting Cyclotron Laboratory, Michigan State University, East Lansing, MI 48824, USA}
\affiliation{Department of Physics and Astronomy, Michigan State University, East Lansing, MI 48824, USA}
\affiliation{Joint Institute for Nuclear Astrophysics, Michigan State University, East Lansing, MI 48824, USA}
\author{S. Gupta}
\altaffiliation{Present address: Indian Institute of Technology Ropar, Rupnagar, Punjab, INDIA 140001}
\affiliation{Theoretical Division, Los Alamos National Laboratory, Los Alamos, NM 87545}
\author{G.W. Hitt}
\author{O. Fawwaz}
\affiliation{Department of Applied Mathematics and Sciences, Khalifa University of Science, Technology, and Research, P.O. Box 127788 Abu Dhabi, UAE}



\date{\today}

\begin{abstract}
\begin{description}
\item[Background] Electron-capture reaction rates on medium-heavy nuclei are an important ingredient for modeling the late evolution of stars that become core-collapse or thermonuclear supernovae. The estimation of these rates requires the knowledge of Gamow-Teller strength distributions in the $\beta^{+}$ direction. Astrophysical models rely on electron-capture rate tables largely based on theoretical models, which must be validated and tested against experimental results.
\item[Purpose] This paper presents a systematic evaluation of the ability of theoretical models to reproduce experimental Gamow-Teller transition strength distributions measured via ($n$,$p$)-type charge-exchange reactions at intermediate beam energies. The focus is on transitions from stable nuclei in the $pf$ shell ($45\leq A \leq64$). In addition, the impact of deviations between experimental and theoretical Gamow-Teller strength distributions on derived stellar electron-capture rates is investigated.
\item[Method] Data on Gamow-Teller transitions from 13 nuclei in the $pf$ shell measured via charge-exchange reactions and  supplemented with results from $\beta$-decay experiments where available, were compiled and compared with strength distributions calculated in shell-models (using the GXPF1a and KB3G effective interactions) and quasi-particle random-phase approximation (QRPA) using ground-state deformation parameters and masses from the finite-range droplet model. Electron-capture rates at relevant stellar temperatures and densities were derived for all distributions and compared.
\item[Results] With few exceptions, shell-model calculations in the $pf$ model space with the KB3G and GXPF1a interactions qualitatively reproduce experimental Gamow-Teller strength distributions of 13 stable isotopes with $45\leq A \leq 64$. Results from QRPA calculations exhibit much larger deviations from the data and overestimate the total experimental Gamow-Teller strengths. For stellar densities in excess of $10^{7}$ g/cm$^{3}$, ground-state electron-capture rates derived from the shell-model calculations using the KB3G (GXPF1a) interaction deviate on average less than 47\% (31\%) from those derived from experimental data for which the location of daughter states at low excitation energies are well established. For electron-capture rates derived from Gamow-Teller strengths calculated in QRPA, the deviations are much larger, especially at low stellar densities.
\item[Conclusions] Based on the limited set of test cases available for nuclei in the $pf$ shell, shell-models using the GXPF1a and KB3G interactions can be used to estimate electron-capture rates for astrophysical purposes with relatively good accuracy. Measures of the uncertainties in these rates can serve as input for sensitivity studies in stellar evolution models. Ground-state electron-capture rates based on the QRPA formalism discussed in the paper exhibit much larger deviations than those based on the shell-model calculations and should be used with caution, especially at low stellar densities.
\end{description}
\end{abstract}


\maketitle


\section{Introduction \label{intro}}

Supernovae play a critical role in the universe. They leave behind neutron stars and black holes and their shockwaves drive mixing of interstellar material and stimulate galactic chemical evolution. Supernovae are also major sources of nucleosynthesis. However, for the two main types of supernovae, core-collapse (type II) and thermonuclear (type Ia) supernovae, the driving mechanisms are still not well understood. A variety of nuclear physics processes play important roles in both types and  improving the nuclear physics input for astrophysical simulations of supernovae has been a strong motivation for a wide variety of nuclear experiments and theoretical calculations.

Electron-captures (EC) on medium-heavy nuclei play an important role in both types of supernovae \cite{LAN03}. During the pre-explosion evolution of core-collapse supernovae, when the Fermi energy of the degenerate electron gas becomes sufficiently high to overcome $Q$-value limitations that restrict EC under terrestrial conditions, the nuclear matter in the stellar core is neutronized and the electron abundance ($Y_{e}$) is reduced by EC reactions \cite{A_bbal79}. Consequently, the electron pressure is reduced, and the energy and entropy drop due to the emission of neutrinos in the EC reactions \cite{heg01,A_hix03,LAN03}. In the early pre-explosion evolution, EC on nuclei in the $pf$ shell are key. Just prior to the collapse, EC on nuclei in the $sdg$ shell also plays an important role.

In type Ia supernovae, which are thought to occur in binary systems in which a white dwarf accretes material from a companion star, the thermonuclear flame leaves behind an equilibrium distribution of iron-region ($pf$-shell) nuclei. Electron-captures on these nuclei reduce the pressure and retard the expansion of the star. EC reactions also reduce the amount of $^{56}$Ni produced, making the explosion less bright (see e.g. \cite{A_bra00}, and references therein). Type Ia supernovae are thought to produce about half of the iron-group nuclei in the solar system. Hence, the nature of their progenitors and the explosion models can be constrained by the condition that these supernovae should not generate amounts of these elements in excess of a factor of two compared to solar system abundances \cite{IWA99}. However, to reliably do so, it is key to use accurate weak-reaction rates in the simulations \cite{A_bra00}.

To establish an accurate database of weak reaction rates of importance for astrophysical simulations, reliable Gamow-Teller (GT) transition-strength distributions must be available for a large number of nuclei. Additionally, EC in stellar environments occurs at relatively high stellar temperatures and transitions from thermally populated excited states can contribute significantly and must also be taken into account. It will, therefore, be impossible to measure a large fraction of the relevant GT strength distributions and one has to rely on theoretical estimates.

In the past, astrophysical simulations often used EC rates calculated by Fuller, Fowler and Newman (FFN) \cite{A_ffn80,A_ffn82a,A_ffn82b,A_ffn85}. These rates were based on GT strengths calculated in an independent particle model (IPM), supplemented with experimental strengths from $\beta$--decay experiments. Based on results from charge-exchange experiments (see e.g. Ref. \cite{OST92}), it has become evident that GT transition strengths are strongly fragmented and quenched compared to the IPM calculations, largely because of the residual interactions between the nuclear constituents. Modern calculations take into account such residual interactions and updated libraries for weak-reaction rates based on shell-model calculations have been produced for nuclei in the $sd$ \cite{ODA94} and $pf$ shells \cite{LAN01a}. For nuclei beyond the $pf$ shell, schemes beyond the shell-model are required (see e.g. \cite{LAN03}, and references therein, for an overview).

In this work, we focus on the accuracy of estimates for EC rates on a variety of $pf$-shell nuclei ($45\leq A \leq 64$) based on theoretical GT strength distributions. One can measure GT strength distributions directly through $\beta$-decay experiments, but the excitation-energy range that can be covered is severely limited by the available $Q$-value window. Therefore, charge-exchange reactions at intermediate energies ($\sim 100$ MeV/$u$ and above) have become the preferred tool to probe GT strength distributions over the full range of excitation energies of interest. The charge-exchange experiments rely on the proportionality between GT transition strengths and the measured differential cross sections extrapolated to values at vanishing linear momentum transfer (\cite{TAD87}; see also Section \ref{SECbgt}). For testing theoretical calculations of GT strength distributions in $pf$-shell nuclei of relevance for EC rates in late stellar evolution, measurements have been performed using the ($n$,$p$), ($d$,$^{2}$He) and ($t$,$^{3}$He) reactions.

Although comparisons of GT strengths and EC rates derived from data and theory have been made for specific nuclei or for subgroups of nuclei with specific theoretical models in the past, a comprehensive evaluation for nuclei in the $pf$ shell is lacking. Such an evaluation is important to gain better insight in the deficiencies of theoretical models, to obtain a measure of the uncertainties in the EC rates used in astrophysical simulations, and to provide guidance to future experimental and theoretical work.
The goals of the current work are two-fold: (i) to perform a consistent comparison of the experimentally extracted GT strength distribution from charge-exchange reactions on stable $pf$-shell nuclei with results from various theoretical models, and (ii) to study the impact of differences between the theoretical and experimental strength distributions on the estimates for EC rates used in astrophysical modeling. Besides the data from the above-mentioned $\Delta T_{z}=+1$ charge-exchange probes on stable $pf$-shell nuclei, we also include data from $\Delta T_{z}=-1$ ($p$,$n$) charge-exchange experiments. Relying on the fact that isospin-symmetry breaking effects are small, they can also be used to extract GT strength distributions of relevance for estimating EC rates. Where available, we also included $\beta$--decay data. In total, 25 distinct data sets were used, for 13 separate nuclei. These nuclei and the data set type(s) are listed in Table \ref{tab:summary}.

Three sets of theoretical calculations have been tested against experiment in the present work. Two of those sets are based on shell-model calculations in the $pf$ model space. These shell-model calculations were performed with the code \textsc{NuShellX@MSU} \cite{NSX,A_nusmsu}. One set of calculations employed the GXPF1a interaction \cite{A_hon02,A_hon04,HON05a}, the other the KB3G interaction \cite{POV01}. The latter interaction is an updated version of the KBF interaction \cite{A_cau99}, which was used to generate the weak reaction rate library of Ref. \cite{LAN01a} and whose parameters were primarily deduced from experimental data in the lower $pf$ shell.
Parameters of the GXPF1a interaction have been fitted to reproduce the experimental excitation energies and masses for many $pf$-shell nuclei.

The third set of theoretical GT strength distributions was based on the Quasi-particle Random Phase Approximation (QRPA) formalism of Ref. \cite{MOL90}, using ground-state deformation parameters and masses from the finite-range droplet model of Ref. \cite{MOL95}. GT strengths produced in this model have been used, for example, to estimate heating in the accreted neutron star ocean \cite{GUP07}. In the following, we refer to this set of calculations as ``QRPA". Although not the focus of the current work, usage of the this QRPA formalism for calculating GT strength distributions has the advantage that it can be applied to nuclei beyond the $pf$-shell.

This paper is organized as follows. In Section \ref{SECbgt}, we briefly review the extraction of GT strengths from the data and the manner in which the experimental and theoretical GT strength distributions were entered in the EC-rate calculation. The EC rate calculations are also briefly described, focusing on input parameters related to the GT strength distributions. In Section \ref{sec:comparison}, a case-by-case and self-contained comparison of experimental and theoretical GT strengths is provided for each of the nuclei listed in Table \ref{tab:summary}. In addition, derived EC rates are also calculated for different astrophysical conditions. In Section \ref{SECdiscussion}, the results from the comparisons between theory and experiments for individual cases are combined in an attempt to gain a better insight into the overall capability of the theoretical models to accurately estimate EC rates of astrophysical interest.

\begin{table*}
\caption{\label{tab:summary}Overview of the GT transitions studied in this work. For each of the cases, the data types available and the types of theoretical calculations used are indicated. In addition, the ground-state to ground-state $Q$-value for electron-capture is provided. The last column gives the references of the papers from which the data sets used in the comparisons to theory were drawn.}
\begin{ruledtabular}
\begin{tabular}{cccccccccccc}
    $i$   & $f$   & $\beta-$decay & ($n$,$p$) & ($d$,$^{2}$He) & ($t$,$^{3}$He) & ($p$,$n$)\footnotemark[1] & QRPA  & KB3G  & GXPF1a & Q$_{EC}$(MeV)\footnotemark[3] & Ref.\\
    \hline
    $^{45}$Sc($\frac{7}{2}^{-}$) & $^{45}$Ca($\frac{5}{2}^{-}$,$\frac{7}{2}^{-}$,$\frac{9}{2}^{-}$) & x     & x     &       &       &       & x     & x     & x & -0.7677 & \cite{FREE65,ALF91}\\
    $^{48}$Ti($0^{+}$) & $^{48}$Sc($1^{+}$) &       & x     & x     &       &       & x     & x     & x & -4.505 & \cite{RAK04,YAK09}\\
    $^{50}$V($6^{+}$) & $^{50}$Ti($5^{+}$,$6^{+}$,$7^{+}$) &       &       & x     &       &       & x     & x     & x & 1.697 & \cite{BAU05}\\
    $^{51}$V($\frac{7}{2}^{-}$) & $^{51}$Ti($\frac{5}{2}^{-}$,$\frac{7}{2}^{-}$,$\frac{9}{2}^{-}$) &       & x     & x     &       &       & x     & x     & x & -2.982 & \cite{ALF93,BAU03} \\
    $^{54}$Fe($0^{+}$) & $^{54}$Mn($1^{+}$) &       & x     &       &       &       & x     & x     & x & -1.2082 & \cite{VET87,VET89} \\
    $^{55}$Mn($\frac{5}{2}^{-}$) & $^{55}$Cr($\frac{3}{2}^{-}$,$\frac{5}{2}^{-}$,$\frac{7}{2}^{-}$) & x     & x     &       &       &       & x     & x     & x & -3.114 & \cite{ELK94,ZOL70,HIL70} \\
    $^{56}$Fe($0^{+}$) & $^{56}$Mn($1^{+}$) &       & x     &       &       &       & x     & x     & x & -4.2072 & \cite{ELK94} \\
    $^{58}$Ni($0^{+}$) & $^{58}$Co($1^{+}$) &       & x     & x     & x     &       & x     & x\footnotemark[2]     & x\footnotemark[2] & -0.893 &\cite{ELK94,HAG04,HAG05,COL06} \\
    $^{59}$Co($\frac{7}{2}^{-}$) & $^{59}$Fe($\frac{5}{2}^{-}$,$\frac{7}{2}^{-}$,$\frac{9}{2}^{-}$) &       & x     &       &       &       & x     & x     & x & -2.0758 & \cite{ALF93} \\
    $^{60}$Ni($0^{+}$) & $^{60}$Co($1^{+}$) &       & x     &       &       & x     & x     & x\footnotemark[2]     & x\footnotemark[2] & -3.334 & \cite{WIL95,ANA08}\\
    $^{62}$Ni($0^{+}$) & $^{62}$Co($1^{+}$) &       & x     &       &       & x     & x     & x     & x & -5.826 & \cite{WIL95,ANA08} \\
    $^{64}$Ni($0^{+}$) & $^{64}$Co($1^{+}$) & x     & x     & x     &       &       & x     & x     & x & -7.818 & \cite{RAH74,WIL95,POP07,ANA08,RAH74} \\
    $^{64}$Zn($0^{+}$) & $^{64}$Cu($1^{+}$) & x     &       & x     & x     &       & x     & x     & x & -1.09 & \cite{SIN07,GRE08,HIT09}\\
\end{tabular}
\end{ruledtabular}
\footnotetext[1]{Using $T_{>}$ transitions and applying isospin symmetry (see text).}
\footnotetext[2]{Shell-model calculations were performed in truncated model space (see text).}
\footnotetext[3]{$Q_{EC}$ is calculated using nuclear masses.}
\end{table*}

\section{GT strengths and electron-capture rates \label{SECbgt}}

If data are available for the decay half-life, GT transition strengths ($B(GT)$) can be calculated directly from the comparative half-life $ft$ using:
\begin{equation}\label{eq:logft}
(\frac{g_{A}}{g_{V}})^{2}B(GT)=\frac{K/g_{V}^2}{ft},
\end{equation}
where $\frac{g_{A}}{g_{V}}=-1.2694\pm 0.0028$ \cite{PDG10} and $K/g_{V}^2=6143\pm2$ s \cite{HAR09}. Here, $B(GT)$ is defined such that it equals 3 for the decay of the free neutron. Of the cases studied here (see Table \ref{tab:summary}), $\beta-$decay data were only available for four nuclei and only provided information about the transition between the ground states of the mother and daughter nuclei.

As mentioned above, the extraction of GT strengths from intermediate-energy ($E\gtrsim 100$ MeV/$u$) charge-exchange reactions relies on the proportionality between the transition strength and the measured differential cross section, extrapolated to zero linear momentum transfer ($q=0$) \cite{TAD87}:
\begin{equation} \label{eq:dsigma}
\left[\frac{d \sigma}{d \Omega}(q=0)\right]_{GT} =\hat{\sigma} B(GT),
\end{equation}
where $\hat{\sigma}$ is the unit cross section. The unit cross section can be derived from direct comparison of the charge-exchange cross section to the $B(GT)$ extracted from $\beta$--decay data, if the latter are available. Otherwise, one relies on the unit cross section extracted from a nucleus of similar mass number, or an empirically established correlation between mass number and unit cross section \cite{TAD87,GRE04,ZEG07,SAS09,PER11}.

When the excitation-energy resolution in the charge-exchange experiments is high compared to the level density, GT strengths can be extracted for each transition to a specific final state. In general, however, this in not the case for the experimental data considered here, or is only true at low excitation energies. Therefore, the GT contributions to the excitation-energy spectra were extracted using a Multipole-Decomposition Analysis (MDA); the $\Delta L=0$, $\Delta S=1$ GT component of the total response in a given excitation-energy bin or peak was determined by fitting the angular distribution to a linear sum of theoretically calculated angular distributions, each associated with different angular momentum transfer. Consequently, the exact location of the GT strength is not necessarily known accurately due to the limited experimental excitation-energy resolution and the analysis of the data into excitation-energy bins of finite width. When preparing the GT transition strengths for the calculation of EC rates, we used the same energy bin widths as used in the original analyses of the data if transitions to specific final states could not be isolated and placed all strengths in the center of the bins. The GT strength distributions are also represented in this manner in the figures. Exceptions were made for the cases where, due to the limited energy resolution, GT strength appeared below the ground state of the final nucleus. Details are provided on a case-by-case basis in Section \ref{sec:comparison}.

Besides the uncertainty in the unit cross section (approximately 10\% \cite{TAD87,ZEG07,SAS09,PER11}), typical uncertainties in the extraction of the GT strength for a transition to a particular final state are $10-20$\%, with error bars increasing for weaker transitions strengths \cite{TAD87,ZEG06,HIT09}. The dominant source of these uncertainties is related to the interference between $\Delta L=0$ and $\Delta L=2$ amplitudes which both can contribute to the $\Delta J=1$ GT transitions. The interference between these amplitudes is caused by the tensor component of the effective nucleon-nucleon interaction that mediates the charge-exchange reaction \cite{LOV81,OST92}. Since the interference can be constructive as well as destructive, depending on the exact nature of the wave functions of the initial and final states, these uncertainties in the extraction of the strength for specific states tend to cancel when summing over several states \cite{ZEG06,HIT09}. The uncertainties in the magnitudes of the extracted GT strengths are not explicitly referred to in the remainder of this paper. Since EC rates are proportional to GT strengths (see below), the uncertainties in the GT strengths extracted from the data limit a meaningful comparison between EC rates based on theoretical and experimental GT strength distribution to a 10-20\% level of discrepancy. As discussed below, the EC rate is much more sensitive to the excitation energies of the states in the daughter nuclei to which EC takes place.

In the shell-model calculations of the GT strengths of relevance for this work, transitions to about 100 final states of a given spin-parity were generated, which was sufficient to cover the excitation energy ranges for which experimental data were available and to ensure that the great majority of the total strength (always more than 95\%) was included in the theoretical results. Except for two cases (detailed below) for which the computational time would have been excessive, the shell-model calculations were performed in the full $pf$ shell, i.e. assuming a $^{40}$Ca inert core. The transition to the ground state of the daughter nucleus is not always of GT nature. Therefore, the level scheme of each daughter nucleus at low excitation energies was also generated in the shell-model calculations (separately for each interaction) to ensure a consistent placement of the first level associated with a GT transition from the parent relative to the ground state calculated within the model. No attempt was made to shift the theoretical strength distribution to better match the data since this cannot be done for the great majority of transitions relevant to astrophysical applications for which no data are available and one thus has to rely on theory only.
The calculated strengths in the shell-models were scaled by a quenching factor which was estimated at $(0.74)^{2}$ for nuclei in the $pf$ shell \cite{MAR96}. The quenching factor accounts for degrees of freedom not included in the shell-model calculations. A large fraction of that quenched strength is known to be located at higher excitation energies, due to the mixing between $1p-1h$ and $2p-2h$ configurations (see e.g. Ref. \cite{YAK05} and references therein). Following Ref. \cite{MOL90}, a quenching factor was not applied to the GT strengths calculated in the QRPA framework; these strength distributions were not modified in any fashion prior to the comparison with the data or the calculation of the EC rates.

The EC rate calculations were performed following the formalism described by Fuller, Fowler and Newman \cite{A_ffn80,A_ffn82a,A_ffn82b,A_ffn85}, which was implemented in a code previously used in Refs. \cite{A_rey06,GUP07}. We briefly review the formalism for calculating EC rates of Refs. \cite{A_ffn80,A_ffn82a,A_ffn82b,A_ffn85}, insofar as they are helpful for the further discussion in this paper.

The EC rate ($\lambda$) is calculated with:
\begin{equation}
\label{eq:lambda}
\lambda=\ln{2}\sum_{j}\frac{f_{j}(T,\rho,U_{F})}{(ft)_{j}},
\end{equation}
where the sum runs over all daughter states that can be populated through a GT transition in the EC reaction. Only transitions from the ground state of the parent nucleus are considered in the calculations presented in this paper. The phase space integral $f_j$ depends on the temperature $T$, density $\rho$ and electrochemical potential $U_{F}$. It is calculated by:
\begin{equation}
\label{eq:phase}
f_j=\int\limits_{w_l}^{\infty}{w^2\left(\frac{Q_{EC}-E_j}{m_ec^2}+w\right)^2G(Z,w)f_e}\,dw,
\end{equation}
where $Q_{EC}$ is the ground-state to ground-state EC
Q-value (calculated with nuclear masses), $E_j$ is excitation energy of the $j^{th}$ daughter state, $m_{e}c^2$ is the electron rest mass, $w=E_e/m_ec^2$ the total electron energy in units of $m_{e}c^2$,  and $G(Z,w)$ is a factor that accounts for the effects of the Coulomb barrier. The lower limit $w_{l}$ corresponds to the threshold value of $w$ for which EC becomes energetically possible.
Realizing that the blocking in the final neutrino phase space plays a negligible role, and neglecting any effects related to the presence of bound electrons and ions, $f_e$ represents the Fermi-Dirac distribution for the energies of the electrons in the stellar plasma:
\begin{equation}
\label{eq:fd}
f_e=\Bigl(\exp\Bigl(\frac{U-U_{F}}{k_{B}T}\Bigr)+1\Bigr)^{-1}.
\end{equation}
$U$ is the kinetic energy of the electron, and $U_{F}/k_{B}T$ is referred to as the degeneracy parameter, and $k_{B}$ is the Boltzmann constant. At $T=0$,
\begin{equation}
\label{eq:potential}
U_{F}(T=0)=0.511\left[\left(1.018(\rho_{6}Y_{e})^{\frac{2}{3}}+1\right)-1\right],
\end{equation}
where $\rho_{6}$ is the density divided by $10^{6}$ g/cm$^{3}$ and $Y_{e}$ is the electron fraction. If the Fermi energy $\varepsilon_{F}=U_{F}+m_{e}c^{2}$ exceeds $Q_{R}=E_{1}-Q_{EC}$, where $E_{1}$ is the excitation energy of the first state that can be captured into in the daughter nucleus,  EC can occur. Eq. (\ref{eq:potential}) shows that with increasing stellar density, capture to daughter states at higher excitation energies becomes increasingly important. With increasing temperature (decreasing degeneracy parameters), the Fermi surface smears out, and EC can occur even when $\varepsilon_{F}($T=0$)<Q_{R}$. If that is the case, the EC rate is particularly sensitive to the excitation energy of the daughter states, and small differences between experimentally and theoretically deduced excitation energies can lead to very large differences in derived EC rates. Therefore, the comparison between EC rates derived from data and theory is difficult if the experimental excitation-energy resolution is poor. Similarly, the ability of a particular theoretical model to accurately predict the excitation energies of the daughter states is critical for the estimation of accurate EC rates.

For the current work, EC rates were calculated over a wide stellar density-temperature grid, where $T$ varied between $10^{7}$ K to $10^{11}$ K and $\rho Y_{e}$ varied between 10 and $10^{14}$ g/cm$^{3}$. However, in the following, we focus on two particular densities: $\rho Y_{e}=10^{7}$ g/cm$^{3}$ (case I) and $10^{9}$ g/cm$^{3}$ (case II), which roughly delineate the density range of interest for EC rates in supernovae. Case I (at $T\approx 3{\times}10^9$ K) is representative for the conditions of the core during the silicon-burning phase of a pre-core-collape progenitor star \cite{A_heg01}. Case II (at $T\approx10{\times}10^9$ K) is representative for the conditions present just prior to the collapse of the core \cite{A_hix03,LAN04}. Case II  is is also representative for the high-density burning regions in which EC occurs during the thermonuclear runaway in Type Ia supernovae \cite{IWA99,A_bra00}.

At $\rho Y_{e}=10^{7} (10^{9})$ g/cm$^{3}$, $\varepsilon_{F}($T=0$)\approx 1.2 (5.2)$ MeV. For many of the nuclei discussed in the paper, $Q_{R}>\varepsilon_{F}$ at the lower of these densities and the EC rates are very sensitive to captures to states at low excitation energies in the daughter nuclei and the temperature. At the higher density, capture into states at higher excitation energies can, depending on the value of $Q_{EC}$, play a significant role.

\section{Comparison of experimental and theoretical GT strength distributions}
\label{sec:comparison}

In this section, we compare the experimental and theoretical GT strength distributions, and the derived EC rates, for each of the cases listed in Table \ref{tab:summary}.

\subsection{$^{45}$Sc$\rightarrow^{45}$Ca \label{SECbgtsc45}}

\begin{figure}
\includegraphics[scale=0.95]{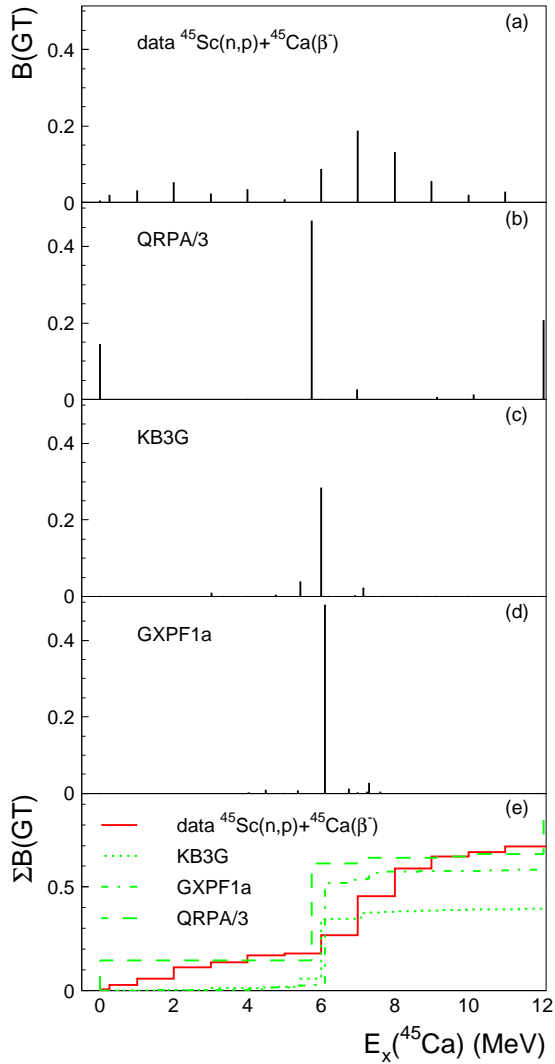}
\caption{\label{bgtsc45} GT transitions from $^{45}$Sc to $^{45}$Ca: (a) from ($n$,$p$) and $\beta-$decay data, (b) from calculations in QRPA (divided by a factor of 3), (c) from shell-model calculation using the KB3G interaction and (d) from shell-model calculations using the GXPF1a interaction. In (e), running sums of $B(GT)$ as a function of excitation energy are plotted. Here, and all similar graphs, the shell-model strengths have been scaled by $(0.74)^{2}$ (see text).}
\end{figure}

\begin{figure}
\includegraphics[scale=0.8]{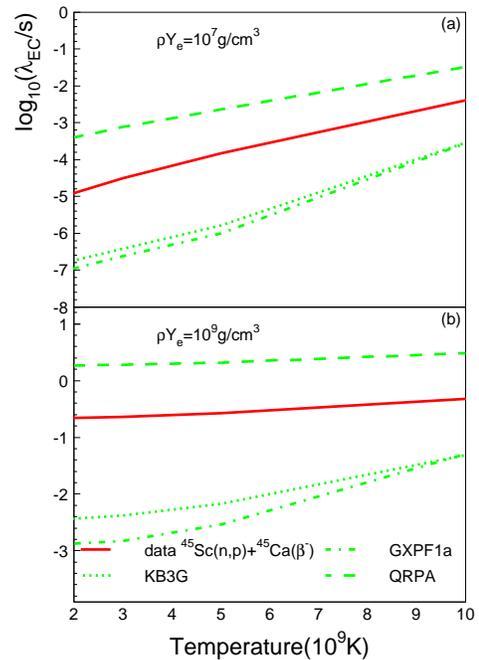}
\caption{Electron-capture rates for $^{45}$Sc($e^-,\nu_e$)$^{45}$Ca based on experimental and theoretical GT strength distribution from Fig. \ref{bgtsc45}, plotted
as a function of temperature at $\rho$Y$_e$=$10^7$ g/cm$^3$ (a)
and $\rho$Y$_e$=$10^9$ g/cm$^3$ (b).\label{ecsc45}}
\end{figure}

Fig. \ref{bgtsc45} shows the experimental and theoretical GT strength distribution for transitions from the $^{45}$Sc($\frac{7}{2}^{-}$) ground state. States in $^{45}$Ca with $J^{\pi}=(\frac{5}{2},\frac{7}{2},\frac{9}{2})^{-}$ are populated. The $B(GT)$ for the transition to the $^{45}$Ca($\frac{7}{2}^{-}$) ground state is known from $\beta$-decay ($4.9\times10^{-3}$) \cite{FREE65}. The GT strength distribution has been measured in a $^{45}$Sc($n,p$) experiment \cite{ALF91}. Because of the limited resolution of that experiment, results are available only in excitation-energy bins of 1 MeV (0.5 MeV for the energy bin between $E_{x}$($^{45}$Ca)=0-0.5 MeV). In Ref. \cite{ALF91} it was noted that, due to the possible contamination from the $(n,p)$ reaction on hydrogen in the target, the extracted strength up to $E_{x}$($^{45}$Ca)=2 MeV is an upper limit. For the present study, the data available from $\beta$-decay and $^{45}$Sc($n,p$) were combined. GT strength in excess of the $\beta-$decay strength reported between $E_{x}$($^{45}$Ca)=0-0.5 MeV in the $^{45}$Sc($n,p$) experiment was placed at $E_{x}$($^{45}$Ca)=0.25 MeV. For $E_{x}$($^{45}$Ca)$>0.5$ MeV, the $B(GT)$ values from the $^{45}$Sc($n,p$) experiment were used, placed at excitation energies in the center of each bin. The combined plot is shown in Fig. \ref{bgtsc45}(a).

Fig. \ref{bgtsc45}(b) shows the results from the QRPA calculation, divided by a factor of 3. Such rescaling of the QRPA strength distributions (for visualization purposes only) is performed throughout the paper.  Figs. \ref{bgtsc45}(c) and (d) show the results from the shell-model calculations in the full $pf$ shell-model space using the KB3G and GXPF1a interactions, respectively. Fig. \ref{bgtsc45}(e) shows the running sum of B(GT) as a function of excitation energy for the data and each of the theoretical calculations.

Under the assumption that the broad distribution seen around 7 MeV in the ($n$,$p$) data is from a transition to a single state in $^{45}$Ca (but broadened due to the limited excitation energy resolution), all three theoretical calculations predict the location of that state about 1 MeV too low. About 0.2 units of GT strength are found below 2 MeV in the experiment. The shell-model calculations using the GXPF1A (KB3G) interaction predict a summed B(GT) of 0.014 (0.033) below $E_{x}$($^{45}$Ca)=5 MeV and a  B(GT) of 0.00023 (0.00019) for the ground-state to ground-state transition, which is about a factor of 10 smaller than the experimental value and hard to distinguish in Figs. \ref{bgtsc45}(c) and (d).  Above 5 MeV, much better consistency is found, although the summed strength up to 10 MeV is better reproduced in the shell-model calculation using the GXPF1a interaction than the calculation using the KB3G interaction. In the QRPA calculations, the ground-state to ground-state transition is about thirty times stronger than deduced from the $\beta$--decay measurement and the total GT strength is overestimated by a factor of 3.

Fig. \ref{ecsc45}(a) and (b) show the EC rates on the ground state of $^{45}$Sc as function of stellar temperature, based on the three theoretical and the experimental GT strength distributions shown in Fig. \ref{bgtsc45}, at values of $\rho Y_{e}$ of $10^7$ g/cm$^{3}$ and $10^9$ g/cm$^{3}$, respectively.
Because $Q_{EC}$=-0.7677 MeV (See Table \ref{tab:summary}), states in $^{45}$Ca at low excitation energies reside below the Fermi surface ($\varepsilon_{F}(T=0)=1.2$ MeV) even at the lower density, and the derived EC rates do not change drastically as a function of temperature. However, the excess of strength seen at low excitation energies in the experimental distribution gives rise to a strongly enhanced ($\times 20-100$) EC rate in comparison to those derived from the shell-model calculations. Conversely,the strong transition to the ground state of $^{45}$Ca in the case of the QRPA calculation results in a much higher rate. At the higher density (Fig. \ref{ecsc45}(b)) the transitions to the excited states start to play a role, including the strong transition to the state at $E_{x}(^{45}$Sc)$\sim 6$ MeV. Nevertheless, since the phase-space is small for a transition at such a high excitation energy the discrepancies between the EC rates seen at the lower density remain at the higher density.
Compared to the other nuclei studied, the case of $^{45}$Sc exhibits amongst the largest discrepancies between the data and shell-model calculations. Early work \cite{SKO74} indicated that mixing between $pf$ and $sd$-shell configurations can occur for states at relatively low excitation energies in the lightest $pf$-shell nuclei, which could affect the GT strength distributions. To confirm and evaluate the impact of possible configuration mixing, further experimental data would be helpful. In addition, it is important to resolve the possible contamination of the experimental GT strength distribution at low-excitation energies from reactions on hydrogen in the $^{45}$Sc target.

\subsection{$^{48}$Ti$\rightarrow^{48}$Sc \label{SECbgtti48}}

\begin{figure}
\includegraphics[scale=0.95]{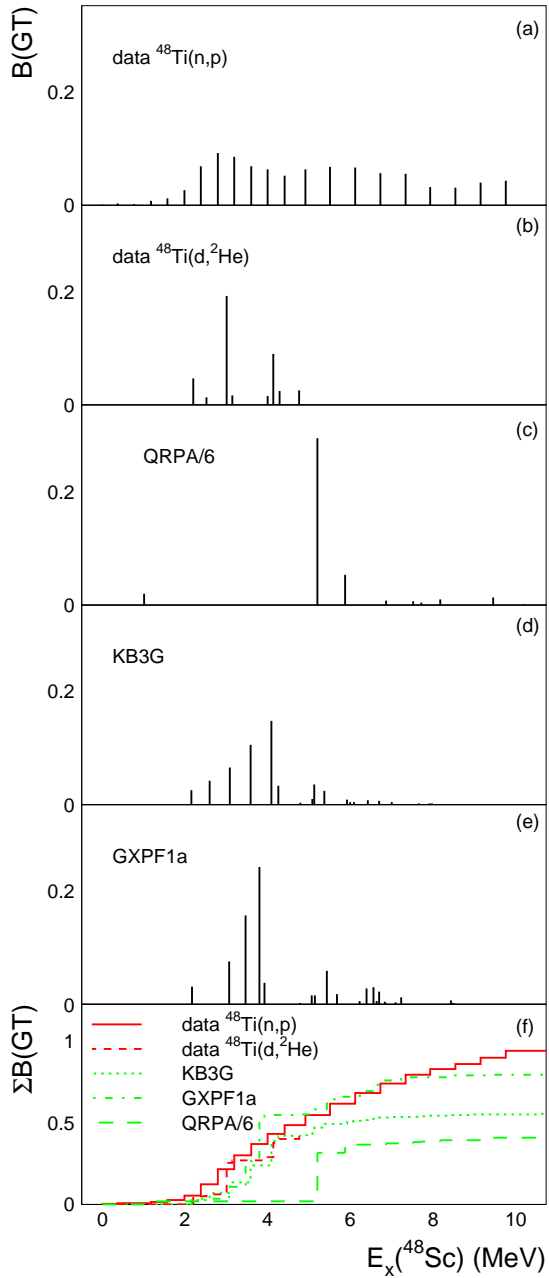}
\caption{\label{bgtti48}GT transitions from $^{48}$Ti to $^{48}$Sc: (a) from ($n$,$p$), (b) from ($d$,$^{2}$He) data, (c) from calculations in QRPA (divided by a factor of 6), (d) from shell-model calculations using the KB3G interaction and (e) from shell-model calculation using the GXPF1a interaction. In (f), running sums of $B(GT)$ as a function of excitation energy are plotted.}
\end{figure}

Fig. \ref{bgtti48} shows the experimental and theoretical GT strength distributions for transitions from the $^{48}$Ti($0^{+}$) ground state to $1^{+}$ final states in $^{48}$Sc. GT strengths from two sets of $^{48}$Ti($n,p)$ data are available. In the earlier experiment \cite{ALF90} performed at $E({n})=200$ MeV, the excitation-energy resolution was 1.2 MeV. In the later experiment \cite{YAK09}, performed at $E({n})=300$ MeV, a similar resolution was achieved, but with better statistical accuracy. In a comparison of the two data sets, we found that below $E_{x}(^{48}$Sc)=6 MeV, the GT strength distributions were consistent, but that at higher excitation energies, slightly less strength was reported in Ref. \cite{YAK09}. Since a higher statistical accuracy allows for a more accurate MDA, we used the data (and adopted the same energy binning) from Ref. \cite{YAK09} in our analysis, which is shown in Fig. \ref{bgtti48}(a).

Fig. \ref{bgtti48}(b) shows the GT strength distribution extracted from a $^{48}$Ti($d$,$^{2}$He) experiment at $E(d$)=183 MeV \cite{RAK04}. The excitation-energy resolution was 120 keV, which made it possible to extract the GT strengths on a state-by-state basis up to $E_{x}(^{48}$Sc)=5 MeV. GT strengths were not extracted beyond that energy. Since $Q_{EC}=-4.505$ MeV, this limit didn't affect the EC rate calculations significantly, as the levels populated beyond $E_{x}(^{48}$)Sc=5 MeV are located above the Fermi level ($\varepsilon_{F}(T=0)=5.2$ MeV), even at $\rho Y_{e}=10^9$ g/cm$^{3}$) and a significant amount of strength is present below the Fermi level.

Fig. \ref{bgtti48}(c) shows the results from the QRPA calculation, divided by a factor of 6.  Figs. \ref{bgtsc45}(d) and (e) show the results from the shell-model calculations in the full $pf$ shell-model space using the KB3G and GXPF1a interactions, respectively. Fig. \ref{bgtti48}(f) shows the running sum of $B(GT)$ as a function of excitation energy.

Both shell-model calculations do about equally well in describing the high-resolution ($d$,$^{2}$He) data, but the calculation with the GXPF1a interaction provides a better estimate for the total strength found up to 10 MeV in the ($n$,$p$) data. This is mostly due to the fact that slightly more strength is predicted at low excitation energies in the calculation with the GXPF1a interaction. Neither the GXPF1a or KB3G interaction describe the long tail seen in the strength distribution extracted from the $(n,p)$ data. In the QRPA calculations, the strength distribution is dominated by a single transition to a state at 5 MeV, in clear contradiction to the data. A weaker transition at about 1 MeV is also predicted, but is not seen in the data.

\begin{figure}
\includegraphics[scale=0.8]{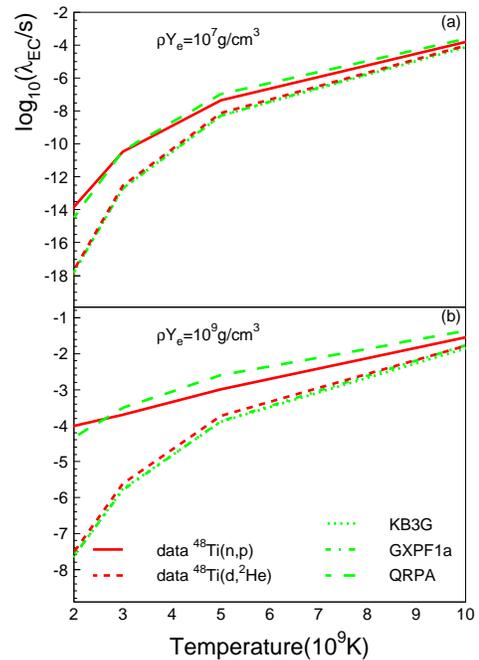}
\caption{Electron-capture rates for $^{48}$Ti($e^-,\nu_e$)$^{48}$Sc, based on experimental and theoretical GT strength distributions shown in Fig. \ref{bgtti48}, plotted as a function of temperature at two densities, $\rho$Y$_e$=$10^7$ g/cm$^3$ (a)
and $\rho$Y$_e$=$10^9$ g/cm$^3$ (b). Note that the rates based on the shell-model calculations with the KB3G and GXPF1a interactions overlap.
\label{ecti48}}
\end{figure}

The EC rates derived from the experimental and theoretical strength distributions is displayed in Figs. \ref{ecti48}. The rates calculated from the strength distributions generated with the GXPF1a and KB3G interactions are similar and match well with those based on the high-resolution $^{48}$Ti($d$,$^{2}$He) data. The rates based on GT strengths extracted from the $^{48}$Ti($n$,$p$) data are much higher, largely due to the poor energy resolution. In combination with the chosen binning of the data, this causes some of the GT strength to be placed at artificially low excitation energies. The rates calculated based on the strength distribution in QRPA are also higher than the rates based on the shell-model calculations and the $^{48}$Ti($d$,$^{2}$He) data due to the relatively strong transition to the above-mentioned state at about 1 MeV. Therefore, the correspondence between the rates based on the QRPA calculations and the $^{48}$Ti($n$,$p$) data is coincidental.

\subsection{$^{50}$V$\rightarrow^{50}$Ti \label{SECbgtv50}}

\begin{figure}
\includegraphics[scale=0.95]{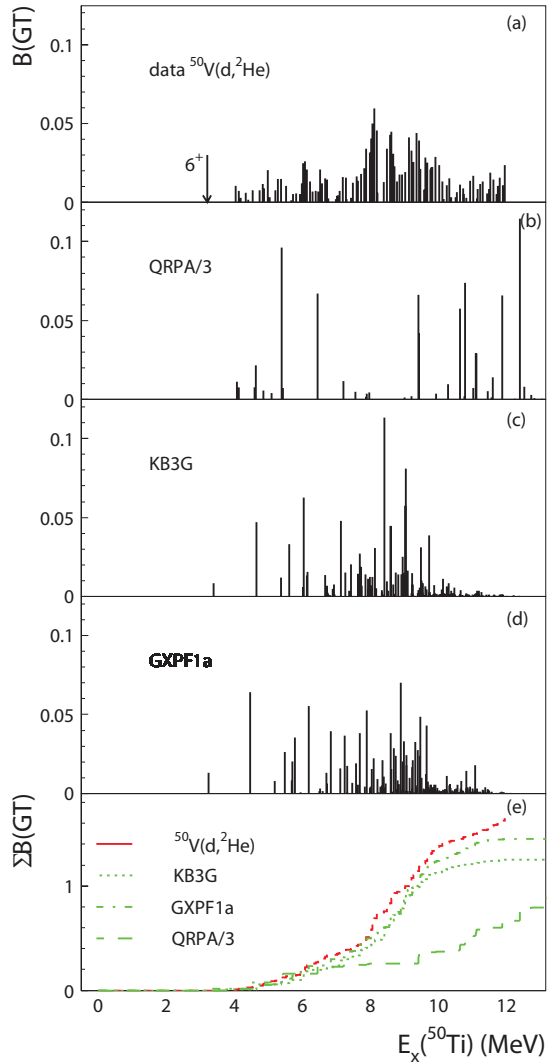}
\caption{\label{bgtv50}GT transitions from $^{50}$V to $^{50}$Ti: (a) from ($d$,$^{2}$He) data (a known but unobserved $6^{+}$ state at 3.198 MeV is indicated), (b) from calculations in QRPA (divided by a factor of 3), (c) from shell-model calculations using the KB3G interaction and (d) from shell-model calculations using the GXPF1a interaction. In (e), running sums of $B(GT)$ as a function of excitation energy are plotted. }
\end{figure}

Information about GT transition strength from the $^{50}$V($6^{+}$) ground state to $(5,6,7)^{+}$ states in $^{50}$Ti are only available from a $^{50}$V($d$,$^{2}$He) experiment performed at 171 MeV \cite{BAU05}. The excitation-energy resolution was 160 keV. The extracted GT strength distribution is shown in Fig. \ref{bgtv50}(a). A known $6^{+}$ state at 3.198 MeV \cite{BUR95}, was obscured by a strong peak from the H($d$,$^{2}$He) reaction at forward scattering angles, but the authors could determine that its $B(GT)$ was "insignificant" from the data at larger scattering angles. The location of this state is indicated in Fig. \ref{bgtv50}(a). Since the high level density made it impossible to assign GT strengths on a state-by-state basis, the strengths shown are in 50-keV wide bins, following Fig. 4(a) from Ref. \cite{BAU05}.

Figs. \ref{bgtv50}(b,c,d) shows the theoretical calculations for the GT strength distribution from the QRPA (divided by 3) and shell-model calculations in full $pf$ model space using the KB3G and GXPF1a interactions, respectively. Fig. \ref{bgtv50}(e) shows the running sum of $B(GT)$ as a function of excitation energy.

The shell-model calculations with the GXPF1a and KB3G interactions are very similar and both correspond quite well to the experimental GT strength distribution. In addition, both predict a weak transition to a $6^{+}$ state situated about 1 MeV below transitions to other states, consistent with the experiment. In both shell-model calculations, somewhat less fragmentation of strength compared to the experimental results is predicted up to an excitation energy of 8 MeV. A similar, but enhanced effect is seen over the excitation energy range up to 11 MeV for the calculations performed in QRPA. Compared to the shell-model calculations and the data, the QRPA calculations also predict a relatively large amount of strength at excitation energies above 9 MeV.

\begin{figure}
\includegraphics[scale=0.8]{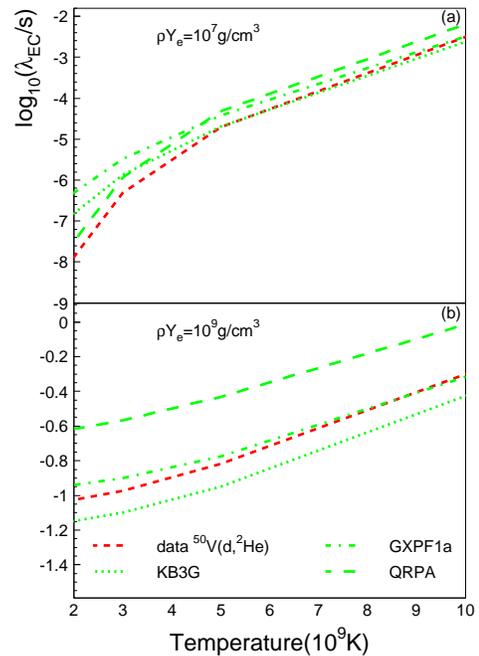}
\caption{Electron-capture rates for $^{50}$V($e^-,\nu_e$)$^{50}$Ti based on experimental and theoretical GT strength shown in Fig. \ref{bgtv50}GT, plotted
as a function of temperature at two densities, $\rho$Y$_e$=$10^7$ g/cm$^3$ (a)
and $\rho$Y$_e$=$10^9$ g/cm$^3$ (b). \label{ecv50}}
\end{figure}

In Figs. \ref{ecv50}(a) and (b), EC rates based on these strengths are displayed. Because the transition to the $6^{+}$ state at 3.198 MeV could not be included in the rate based on the $^{50}$V($d$,$^{2}$He) experiment, the corresponding rate estimate is lower. If an EC transition to this $6^{+}$ state, with a strength similar to that calculated with the KB3G and GXPF1a interactions, was artificially included when generating the EC rates based on the $^{50}$V($d$,$^{2}$He) experiment, the rate at $\rho$Y$_e$=$10^7$ g/cm$^3$ matches about equally well to either of these two theoretical rates. At $\rho$Y$_e$=$10^7$ g/cm$^3$, the calculation that employs the GXPF1a interaction does slightly better, independent of whether a small contribution from transitions to the first $6^{+}$ state was artificially included for the rate calculation based on the $^{50}$V($d$,$^{2}$He) data.

The value of $Q_{EC}$ for the transition from the $^{50}$V ground state to the $^{50}$Ti ground state is positive, unlike the other cases studied here (see Table \ref{tab:summary}). Combined with the weakness of the transition to the first $6^{+}$ state, the transitions to the higher-lying excited states thus play a significant role and the rate is somewhat less sensitive to the details of the low-lying excitations. Therefore, even though the GT strength distribution calculated in QRPA is quite different from that observed in the experiment, the deduced EC rates coincidentally match quite well.

\subsection{$^{51}$V$\rightarrow^{51}$Ti \label{SECbgtv51}}

\begin{figure}
\includegraphics[scale=0.95]{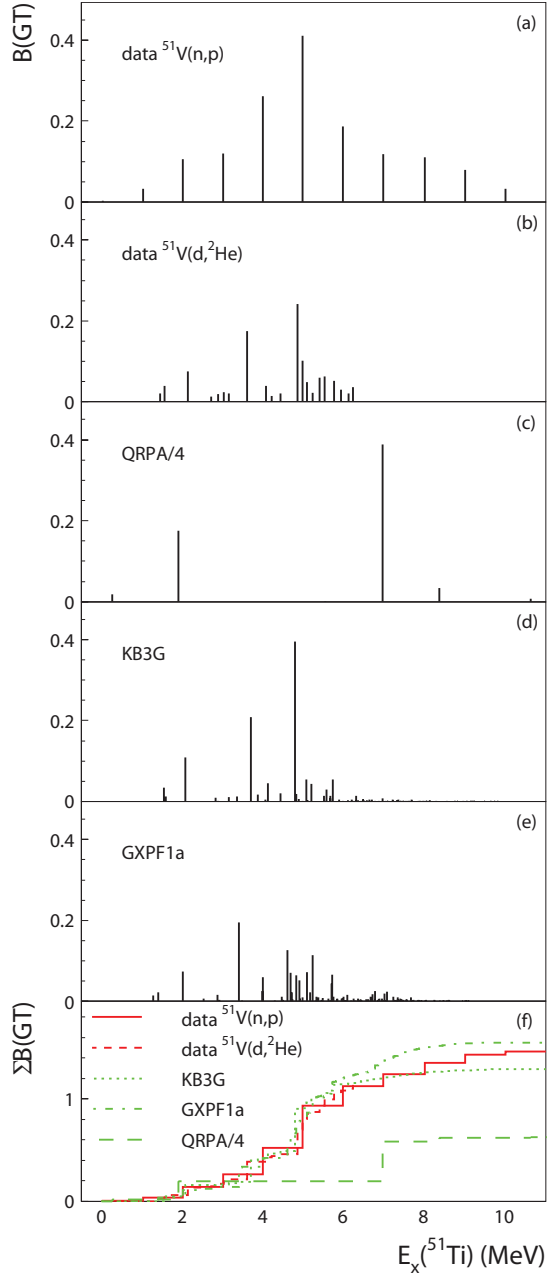}
\caption{\label{bgtv51}GT transitions from $^{51}$V to $^{51}$Ti: (a) from ($n$,$p$) data, (b) from ($d$,$^{2}$He) data, (c) from calculations in QRPA (divided by a factor of 4), (d) from shell-model calculations using the KB3G interaction and (e) from shell-model calculations using the GXPF1a interaction. In (f), running sums of $B(GT)$ as a function of excitation energy are plotted.}
\end{figure}

Experimental information about GT transitions from the $^{51}$V $(\frac{7}{2}^{-}$) ground state to $(\frac{5}{2},\frac{7}{2},\frac{9}{2})^{-}$ final states in $^{51}$Ti is available from a $^{51}$V($n,p$) experiment \cite{ALF93} and a $^{51}$V($d,^{2}$He) experiment \cite{BAU03}.
The excitation-energy resolution achieved in the ($n,p$) experiment was 0.9 MeV and the results are presented in Fig. \ref{bgtv51}(a). Bins of 1-MeV wide were used, as in the original work. The results from the $^{51}$V($d,^{2}$He) experiment are shown in Fig. \ref{bgtv51}(b). A resolution of 110 keV was achieved, allowing for a detailed extraction of $B(GT)$ values up to $E_{x}\approx6.5$ MeV. Here, we used Fig. 3(top) in Ref. \cite{BAU03} to obtain $B(GT)$ values on a state-by-state basis.

Figs. \ref{bgtv51}(c,d,e) show the theoretical GT strength distribution from the QRPA (divided by a factor of 4) and shell-model calculations in full $pf$ model space using the KB3G and GXPF1a interactions, respectively. Fig. \ref{bgtv51}(f) shows the running sum of $B(GT)$ as a function of excitation energy.

The GT strength distributions calculated in the shell-model using the GXPF1a and KB3G interactions agree well with those extracted from the data. Above 6.5 MeV, where ($d,^{2}$He) strengths are not available, a considerable amount of strength is seen in the ($n,p$) data. The shell-model calculations predict some strength to be located at higher excitation energies, but slightly less than observed in the ($n,p$) data. The QRPA strength distribution exhibits little fragmentation and produces two strong GT transitions at $\sim 2$ and $\sim 7$ MeV. A weak transition is predicted at very low excitation energies, but not seen in the high-resolution $^{51}$V($d,^{2}$He) data. The summed GT strength up to 11 MeV in the QRPA calculation is about 40\% higher than seen in the $^{51}$V($n,p$) data.

\begin{figure}
\includegraphics[scale=0.8]{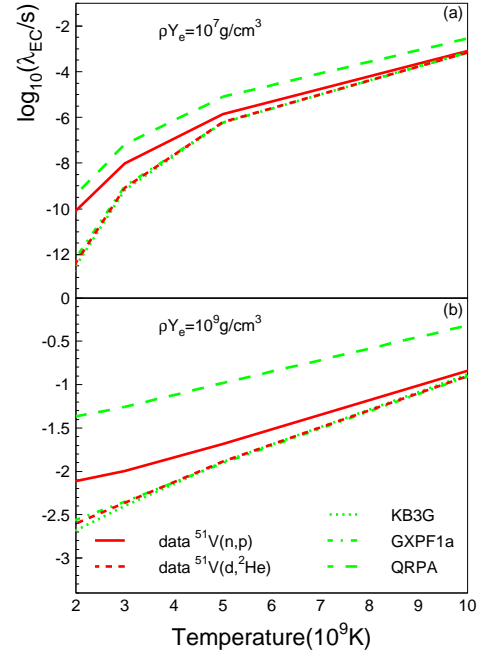}
\caption{Electron-capture rates for $^{51}$V($e^-,\nu_e$)$^{51}$Ti, based on experimental and theoretical GT strength distributions shown in Fig. \ref{bgtv51},  plotted
as a function of temperature at two densities, $\rho$Y$_e$=$10^7$ g/cm$^3$ (a)
and $\rho$Y$_e$=$10^9$ g/cm$^3$ (b). Note that the rates based on the shell-model calculations with the KB3G and GXPF1a interactions overlap. \label{ecv51}}
\end{figure}

The close correspondence between the GT strengths calculated in the shell-model and those extracted from the $^{51}$V($d,^{2}$He) data results in a close match of the deduced EC rates, as shown in Fig. \ref{ecv51}. Since $Q_{EC}=-2.982$ MeV and the bulk of the GT strength is located below 6 MeV, the strength not extracted at high excitation energies in the $^{51}$V($d,^{2}$He) data is of little consequence, even at the higher density.
Because of the limited resolution achieved in the $^{51}$V($n,p$) experiment, some strength is found at relatively low excitation energies, resulting in an increase of the EC rate, in particular at low temperatures. In the QRPA calculation, excess strength is predicted at low excitation energies. Combined with the prediction of a low-lying state not observed in the high-resolution $^{51}$V($d,^{2}$He) data, the EC rates calculated based on the QRPA calculations are significantly higher than those based on that data, at both densities considered.

\subsection{$^{54}$Fe$\rightarrow^{54}$Mn \label{SECbgtfe54}}

\begin{figure}
\includegraphics[scale=0.95]{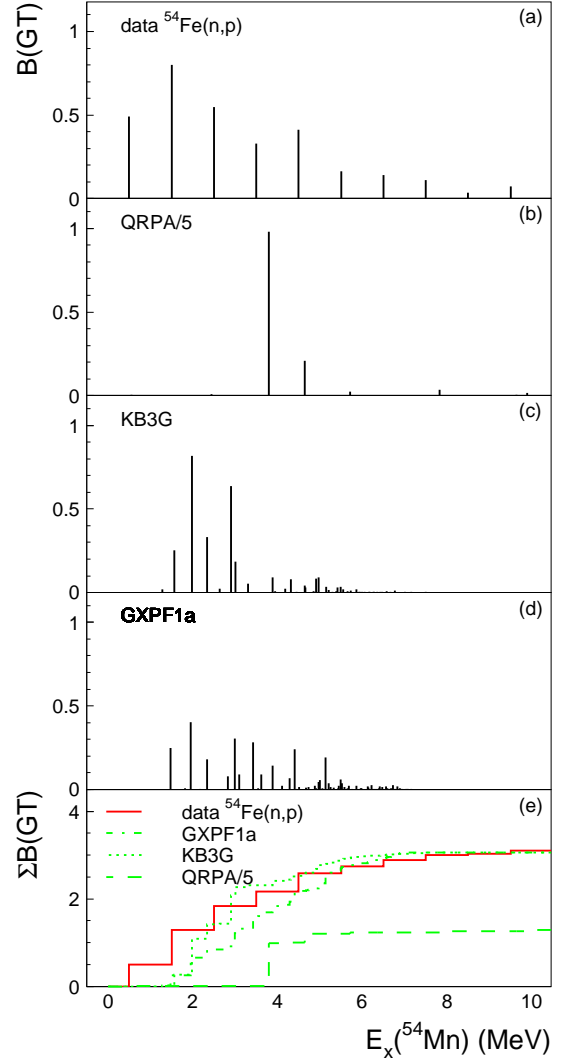}
\caption{\label{bgtfe54}GT transitions from $^{54}$Fe to $^{54}$Mn: (a) from ($n$,$p$) data, (b) from calculations in QRPA (divided by a factor of 5), (c) from shell-model calculations using the KB3G interaction and (d) from shell-model calculations using the GXPF1a interaction. In (e), running sums of $B(GT)$ as a function of excitation energy are plotted.}
\end{figure}

Information about GT transitions from the $^{54}$Fe($0^{+}$) ground state to $1^{+}$ states in $^{54}$Mn is available from a $^{54}$Fe($n,p$) experiment, performed at 300 MeV \cite{VET87,VET89}. The excitation-energy resolution was 1.2 MeV. A second $^{54}$Fe($n,p$) experiment was performed at 100 MeV \cite{RON93}, but since the resolution was relatively poor (2.8 MeV), we used the earlier data for the comparison with the theoretical calculation. Those results are shown in Fig. \ref{bgtfe54}(a). Since Table I in Ref. \cite{VET89} gives the $B(GT)$ values in 2-MeV wide bins, we have instead used Fig. 10(a) and the quoted GT unit cross section of 5.1$\pm$0.8 mb/sr in that reference to produce a GT strength distribution in 1-MeV wide bins.
We note that a $^{54}$Fe($d$,$^{2}$He) experiment has also been performed \cite{AJU01} with slightly better resolution ($\sim700$ keV) than in the above-mentioned $^{54}$Fe($n,p$) experiments. However, the ($n,p$) data was used to extract information about GT excitations from the $^{54}$Fe($d$,$^{2}$He) experiment, which made it impossible to treat the latter as an independent measurement. It is, therefore, not included in the present analysis.

Figs. \ref{bgtfe54}(b,c,d) show the theoretical calculations for the GT strength distribution from the QRPA calculation (divided by a factor of 5) and shell-model calculations in full $pf$ model space using the KB3G and GXPF1a interactions, respectively. Fig. \ref{bgtfe54}(e) shows the running sums of $B(GT)$ as a function of excitation energy.

Both shell-model calculations do well in describing the total GT strength up to an excitation energy of 10 MeV extracted from the $^{54}$Fe($n,p$) data, but the strength distribution is spread out more in the calculation with the GXPF1a interaction than in the calculation with the KB3G interaction. In the $^{54}$Fe($n,p$) data a significant amount of strength is found between 0.5 MeV and 1.5 MeV. In contrast, both shell-model calculations predict the lowest transitions to occur to $1^{+}$ states in $^{54}$Mn at about 1.5 MeV, which is consistent with the location of positively identified $1^{+}$ states at 1.391 MeV and 1.454 MeV \cite{ENSDF}. The results from the $^{54}$Fe($d$,$^{2}$He) experiment \cite{AJU01} also seem to indicate that a large fraction of the strength found below 1.5 MeV in the ($n$,$p$) experiment is in fact due to the transitions to the states located at $E_{x}\approx 1.4$ MeV. We conclude that the correspondence between the shell-model calculations and the data would probably improve if high-resolution spectra were available.
The QRPA calculations do not reproduce the experimental spectrum well, concentrating nearly all GT strength in the population of two states at $E_{x}\approx4$ MeV. The summed strength up to 10 MeV is 60\% higher than observed in the data. Although not distinguishable in Fig. \ref{bgtfe54}(b), a weak ($B(GT)=0.016$) transition to a state at $E_{x}=0.55$ MeV is also predicted.

\begin{figure}
\includegraphics[scale=0.8]{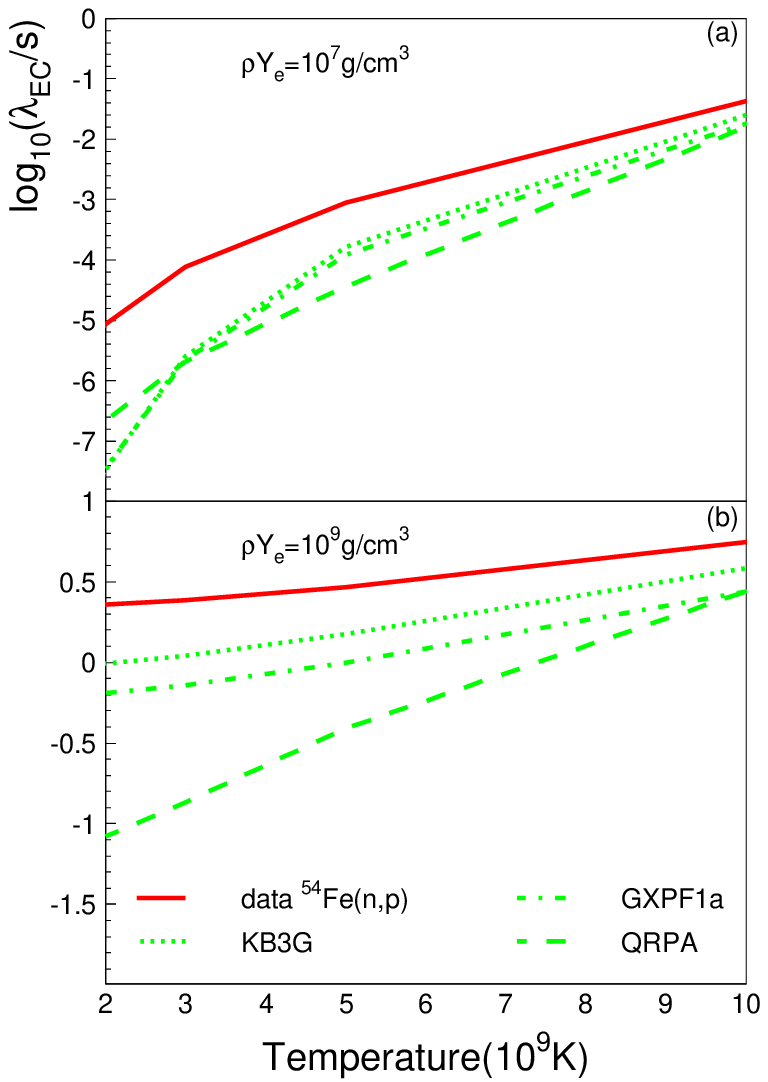}
\caption{Electron-capture rates for $^{54}$Fe($e^-,\nu_e$)$^{54}$Mn based on experimental and theoretical GT strengths shown in Fig. \ref{bgtfe54}. plotted
as a function of temperature at two densities, $\rho$Y$_e$=$10^7$ g/cm$^3$ (a)
and $\rho$Y$_e$=$10^9$ g/cm$^3$ (b).  \label{ecfe54}}
\end{figure}

Given the relatively large amount of GT strength placed at low-excitation energies, the deduced EC rate from the $^{54}$Fe($n,p$) experiment is significantly higher than those deduced from the theoretical strength distributions, as shown in Fig. \ref{ecfe54}. Because of the effects related to the poor resolution of the data discussed above, this discrepancy is at least partially artificial.
Due to the presence of the weak transition to the $1^{+}$ state at 0.55 MeV predicted in the QRPA calculations, the EC rates at low density (Fig. \ref{ecfe54}(a)) are quite similar to those based on the shell-model calculations. However, the lack of any significant GT strength up to 3.7 MeV results in a large underestimation of the EC rate at the higher densities (Fig. \ref{ecfe54}(b)), in particular at low temperatures.

\subsection{$^{55}$Mn$\rightarrow^{55}$Cr \label{SECbgtmn55}}

\begin{figure}
\includegraphics[scale=0.95]{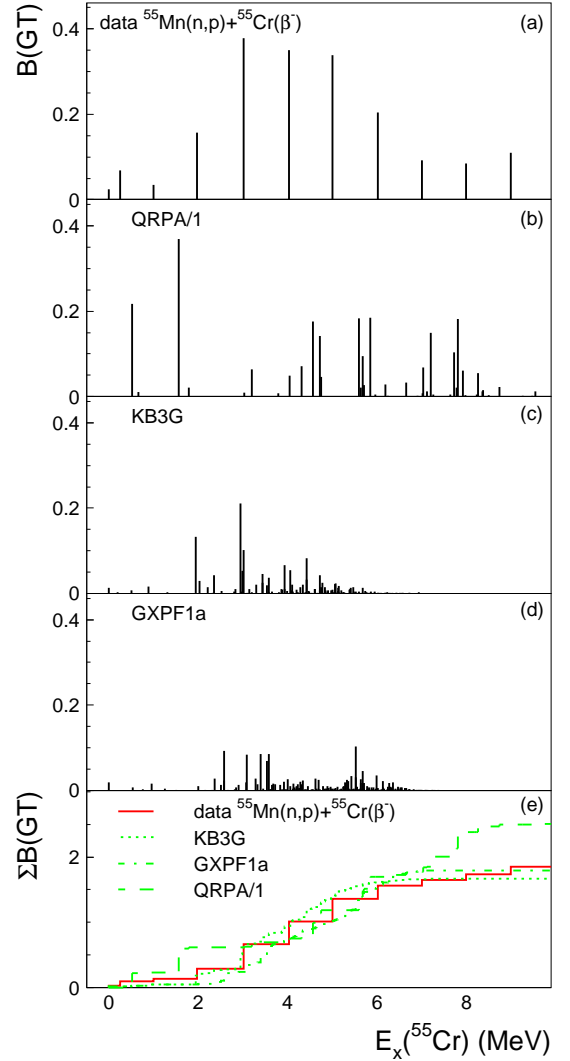}
\caption{\label{bgtmn55}GT transitions from $^{55}$Mn to $^{55}$Cr: (a) from ($n$,$p$) and and $\beta-$decay data, (b) from calculations in QRPA, (c) from shell-model calculations using the KB3G interaction and (d) from shell-model calculations using the GXPF1a interaction. In (e), running sums of $B(GT)$ as a function of excitation energy are plotted.}
\end{figure}

The GT strength distribution for transitions from the $^{55}$Mn$(\frac{5}{2}^{-})$ ground state to $(\frac{3}{2},\frac{5}{2},\frac{7}{2})^{-}$ states in $^{55}$Cr has been measured via the $^{55}$Mn($n,p$) reaction at 198 MeV \cite{ELK94}. Additionally, the $B(GT)$ for the $^{55}$Mn$(\frac{5}{2}^{-}$, g.s.)$\rightarrow^{55}$Cr($\frac{3}{2}^{-}$, g.s.) transition is known from $^{55}$Cr($\beta^{-}$) data \cite{ZOL70,HIL70}. Fig. \ref{bgtmn55}(a) shows the experimental GT distribution, based on a combination of results from the $^{55}$Mn($n,p$) and $^{55}$Cr($\beta^{-}$) experiments. The $B(GT)$ for the transition to the ground state was fixed to the measured value (0.0242) from $^{55}$Cr($\beta^{-}$). Additional strength measured below 0.5 MeV in the $^{55}$Mn($n,p$) experiment was placed at $E_{x}$($^{55}$Cr)=0.25 MeV. For the remainder of the spectrum, values from Ref. \cite{ELK94} were adopted. Figs. \ref{bgtmn55}(b-d) show the theoretical calculations for the GT strength distribution. Fig. \ref{bgtmn55}(e) shows the running sums of $B(GT)$ as a function of excitation energy.

The shell-model calculations using the KB3G and GXPF1a interactions reproduce the experimental spectrum moderately well, including the transition to the ground state of $^{55}$Cr.  In the QRPA calculations, too much strength is predicted below 2 MeV and the total strength up to 10 MeV is about 30\% higher than in the data. In addition, no strength is predicted for the transition to the $^{55}$Cr ground state and the first transition populates a state at 0.51 MeV.

\begin{figure}
\includegraphics[scale=0.8]{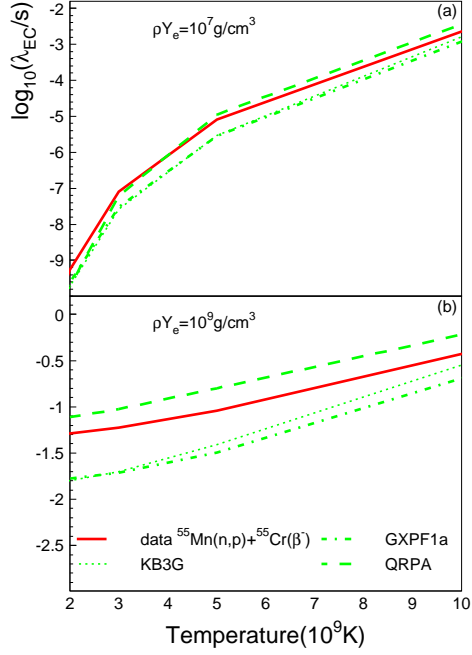}
\caption{Electron-capture rates for $^{55}$Mn($e^-,\nu_e$)$^{55}$Cr based on experimental and theoretical GT strengths shown in Fig. \ref{bgtmn55}, plotted
as a function of temperature at two densities, $\rho$Y$_e$=$10^7$ g/cm$^3$ (a)
and $\rho$Y$_e$=$10^9$ g/cm$^3$ (b).  \label{ecmn55}}
\end{figure}

The EC rates calculated by using the experimental GT strength distribution is slightly higher than the EC rates based on the shell-model calculations (see Fig. \ref{ecmn55}). However, it is difficult to draw very strong conclusions from these minor discrepancies because of the limited resolution of the $(n,p)$ experiment. Since $Q_{EC}=-3.114$ MeV, the calculated EC rates are very sensitive to the strength distributions at low excitation energies, in particular at the lower density (Fig. \ref{ecmn55}(a)). The EC rates based on the QRPA calculations are slightly higher than the rates based on the data and the shell-model calculations (except at the lower density and the lowest temperatures), due to the presence of the two strong transitions below 2 MeV, which are not observed experimentally and do not appear in the shell-model calculations.

\subsection{$^{56}$Fe$\rightarrow^{56}$Co \label{SECbgtfe56}}

\begin{figure}
\includegraphics[scale=0.95]{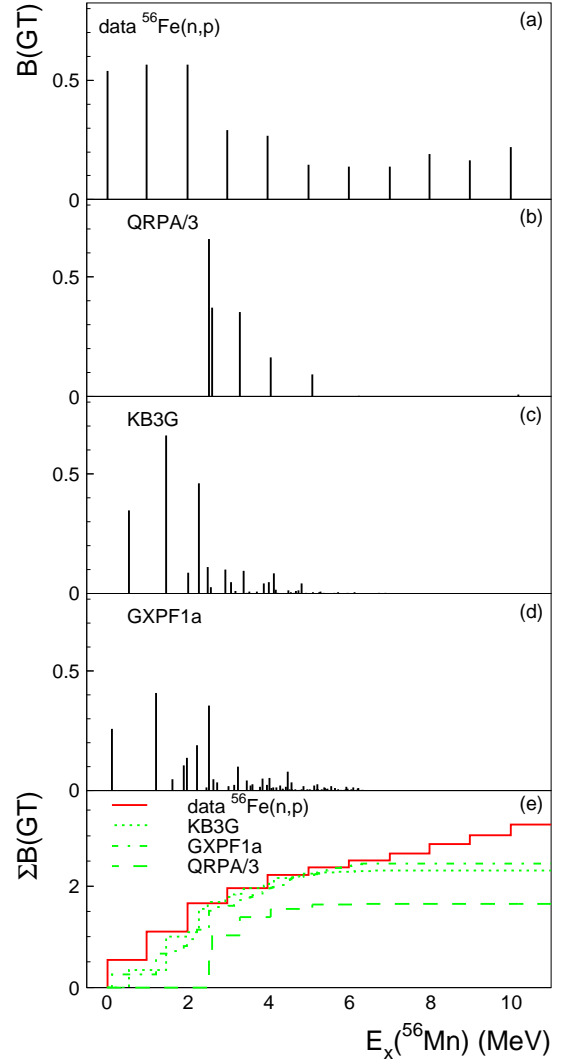}
\caption{\label{bgtfe56}GT transitions from $^{56}$Fe to $^{56}$Mn: (a) from ($n$,$p$) data, (b) from calculations in QRPA (divided by a factor of 3), (c) from shell-model calculations using the KB3G interaction and (d) from shell-model calculations using the GXPF1a interaction. In (e), running sums of $B(GT)$ as a function of excitation energy are plotted.}
\end{figure}

Information about GT transitions from the ground state of $^{56}$Fe($0^{+}$) to $^{56}$Mn is available from a $^{56}$Fe($n,p$) experiment, performed at 198 MeV \cite{ELK94}. The excitation-energy resolution was 1.3 MeV. A second $^{56}$Fe($n,p$) experiment performed at 100 MeV \cite{RON93} had a relatively poor excitation-energy resolution of 2.8 MeV. Hence, we used the earlier data which is shown in Fig. \ref{bgtfe56}(a). In the current analysis, strength reported below $E_{x}=0$ MeV in Ref. \cite{ELK94} was added to the first energy bin. It should be noted that high-resolution (110 keV FWHM) $^{56}$Fe($d$,$^{2}$He) data are available \cite{FRE05}, but GT strengths were not presented in that reference. Figs. \ref{bgtfe56}(b-d) show the theoretical calculations for the GT strength distribution. Fig. \ref{bgtfe56}(e) shows the running sums of $B(GT)$ as a function of excitation energy.

Taking into acount the limited experimental resolution of the $^{56}$Fe($n,p$) data, the shell-model calculations  match well with the data up to an excitation energy of about 7 MeV. Above that energy, significant amounts of GT strength were extracted from the ($n$,$p$) data, whereas no strength is present in the calculated distributions. The QRPA calculations do not reproduce the experimental strength distribution well, missing the strength below 2 MeV and overestimating the total strength by about 50\%.
Although GT strengths are not available from the high resolution $^{56}$Fe($d$,$^{2}$He) experiment \cite{FRE05}, that data set clearly shows two strong transitions to $1^{+}$ states at 0.11 MeV and 1.2 MeV. The shell-model calculations with both interactions exhibit a similar feature, although the calculation with the KB3G interaction locates the first $1^{+}$ state slightly too high in excitation energy.

\begin{figure}
\includegraphics[scale=0.8]{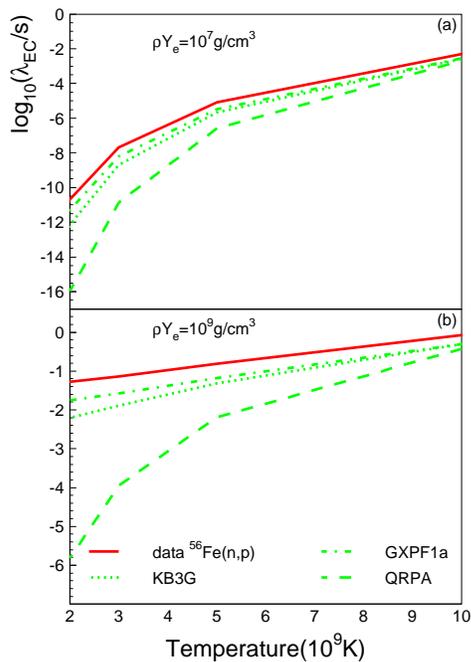}
\caption{Electron-capture rates for $^{56}$Fe($e^-,\nu_e$)$^{56}$Mn based on experimental and theoretical GT strengths plotted in Fig. \ref{bgtfe56},
as a function of temperature at two densities, $\rho$Y$_e$=$10^7$ g/cm$^3$ (a)
and $\rho$Y$_e$=$10^9$ g/cm$^3$ (b).  \label{ecfe56}}
\end{figure}

The EC rates based on the GT strengths calculated in the shell-models fall slightly below those based on the strengths extracted from the $^{56}$Fe($n,p$) as shown in Fig. \ref{ecfe56}. The minor discrepancies are likely enhanced by the poor resolution of the ($n,p$) data, which results in the placement of some GT strength at artificially low excitation energies. Since the excitation energy of the first state is slightly higher in the calculation with the KB3G interaction than with the GXPF1a interaction, the EC rates are slightly lower. Since the excitation energy of the first state that can be captured into is much higher in the QRPA calculations than in the data or the shell-model calculations, the associated EC rates are much lower. At increasing temperatures, when the strengths at higher excitation energies play a more prominent role, the discrepancies reduce, assisted by the fact that the total GT strength predicted in the QRPA calculation is about 50\% higher than in the data or the shell-model calculations.

\subsection{$^{58}$Ni$\rightarrow^{58}$Co \label{SECbgtni58}}

\begin{figure}
\includegraphics[scale=0.95]{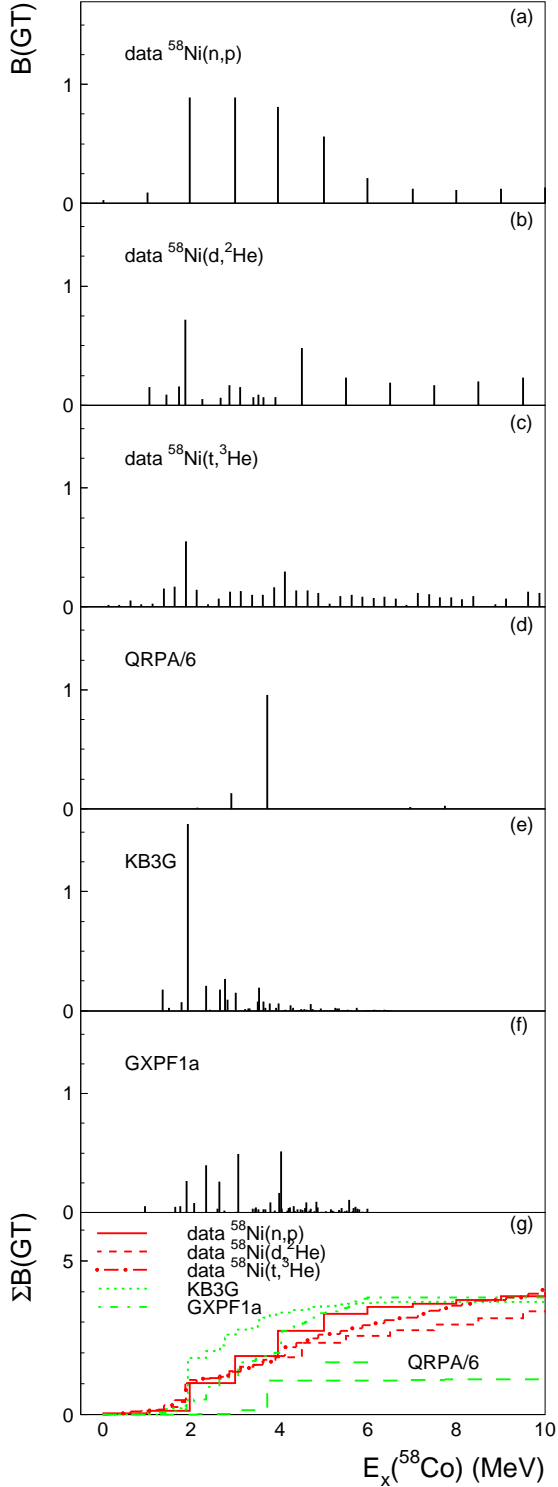}
\caption{\label{bgtni58}GT transitions from $^{58}$Ni to $^{58}$Co: (a) from ($n$,$p$) data, (b) from ($d$,$^{2}$He) data, (c) from ($t$,$^{3}$He) data (d) from calculations in QRPA (divided by a factor of 6), (e) from shell-model calculations using the KB3G interaction and (f) from shell-model calculations using the GXPF1a interaction. In (g), running sums of $B(GT)$ as a function of excitation energy are plotted.}
\end{figure}

Experimental information on GT transitions from the $^{58}$Ni($0^{+}$) ground state to $1^{+}$ states in $^{58}$Co is available from $^{58}$Ni($n,p$) \cite{ELK94}, $^{58}$Ni($d$,$^{2}$He) \cite{HAG04,HAG05} and $^{58}$Ni($t$,$^{3}$He) \cite{COL06} experiments. The results are displayed in Figs. \ref{bgtni58}(a), (b) and (c), respectively. The energy resolutions in the $^{58}$Ni($n,p$), $^{58}$Ni($d$,$^{2}$He) and $^{58}$Ni($t$,$^{3}$He) data sets were 1.2 MeV, 130 keV and 250 keV, respectively. The $^{58}$Ni($n,p$) data was available in energy bins of 1 MeV. In the current analysis, strength recorded below $E_{x}=0$ MeV was added to the first energy bin. For excitation energies below 4 MeV, GT strengths from the $^{58}$Ni($d$,$^{2}$He) experiment were available on a state-by-state basis. Above that excitation energy, an energy  binning of 1 MeV was used. The GT strengths from the $^{58}$Ni($t$,$^{3}$He) experiment were available in 250-keV wide bins up to an excitation energy of 10 MeV.
Figs. \ref{bgtni58}(d-f) show the theoretical calculations for the GT strength distribution from the QRPA (divided by a factor of 6) and shell-model calculations in truncated $pf$ model space (5 holes in the $f_{7/2}$ shell) using the KB3G and GXPF1a interactions, respectively. Fig. \ref{bgtni58}(g) shows the running sums of $B(GT)$ as a function of excitation energy.

The main difference between the shell-model calculations and the experimental data occurs at an excitation energy of about 2 MeV. In the calculation with the KB3G interaction, too much strength is assigned to a single transition, whereas in the calculation with the GXPF1a interaction, too little strength is present near that excitation energy. Since slightly more strength is present at higher excitation energies in the calculation with the GXPF1a interaction, the summed strengths up to 10 MeV for the two shell-model calculations are very close and also consistent with the experimental results. In the QRPA calculations, most of the strength is concentrated around an excitation energy of 4 MeV and the summed strength is a factor of two higher than observed in the data. A very weak transition ($B(GT)=0.00321$) populates a $1^{+}$ state at 0.21 MeV; it is not visible in Fig. \ref{bgtni58}(d).

\begin{figure}
\includegraphics[scale=0.8]{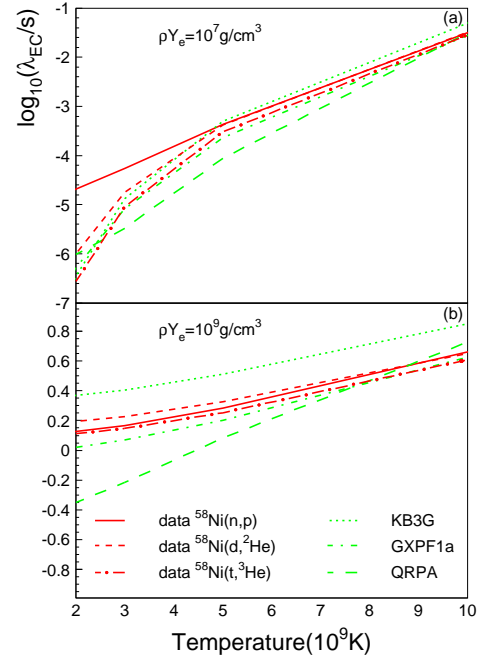}
\caption{Electron-capture rates for $^{58}$Ni($e^-,\nu_e$)$^{58}$Co based on experimental and theoretical GT strengths from Fig. \ref{bgtni58},  plotted
as a function of temperature at two densities, $\rho$Y$_e$=$10^7$ g/cm$^3$ (a)
and $\rho$Y$_e$=$10^9$ g/cm$^3$ (b).
\label{ecni58}}
\end{figure}

The EC rates calculated from the three data sets match each other quite well, as shown in Fig. \ref{ecni58}. Only at the lower density and at low temperatures (Fig. \ref{ecni58}(a)) is the rate based on the $^{58}$Ni($n,p$) data significantly higher than those based on the $^{58}$Ni($d$,$^{2}$He) and $^{58}$Ni($t$,$^{3}$He) data sets. This is likely due to the relatively poor energy resolution achieved in the $^{58}$Ni($n,p$) experiment, which results in the placement of a small amount of strength at excitation energies below 1 MeV.

At $\rho$Y$_e$=$10^7$ g/cm$^3$, the EC rates based on GT strengths predicted in the shell-model calculations do well in reproducing those based on the GT strengths extracted from the three data sets. At the higher density, the EC rate deduced from the calculation with the KB3G interaction is too high, due to the fact that too much strength is placed at relatively low excitation energies. For the EC rate deduced from the calculation with the GXPF1a interaction the reverse is the case, although the discrepancy with the rates based on the experimental data is smaller and vanishes at the highest temperatures. EC rates based on the QRPA calculations are too low, except for the low-temperature region at the lower density (due to the predicted weak transition to a state at 0.21 MeV) and for the high-temperature region at the higher density (where the strong states near 4 MeV start playing a large role).

\subsection{$^{59}$Co$\rightarrow^{59}$Fe \label{SECbgtco59}}

\begin{figure}
\includegraphics[scale=0.95]{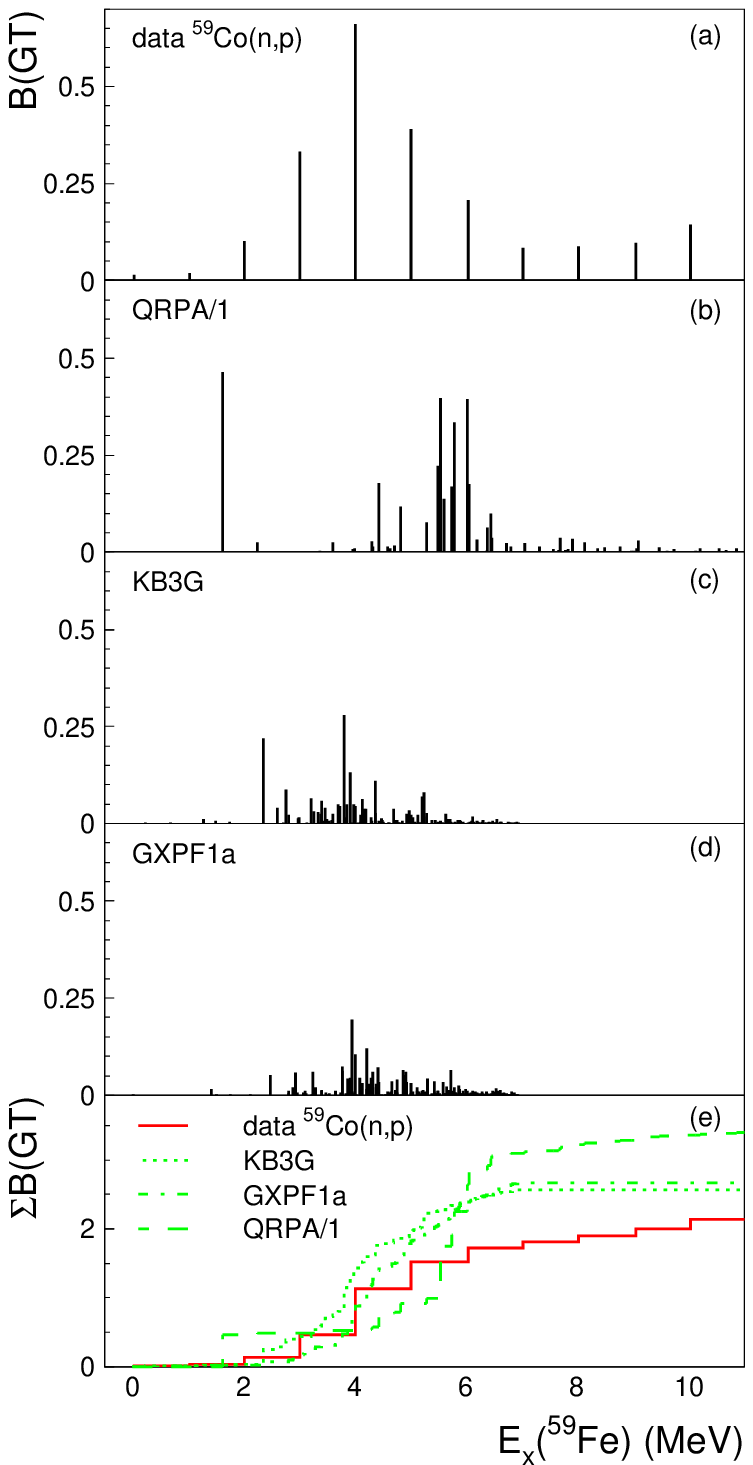}
\caption{\label{bgtco59}GT transitions from $^{59}$Co to $^{59}$Fe: (a) from ($n$,$p$) data, (b) from calculations in QRPA (without additional scaling), (c) from shell-model calculations using the KB3G interaction and (d) from shell-model calculations using the GXPF1a interaction. In (e), running sums of $B(GT)$ as a function of excitation energy are plotted.}
\end{figure}

The GT strength distribution for transitions from the $^{59}$Co$(\frac{7}{2}^{-})$ ground state to $(\frac{5}{2},\frac{7}{2},\frac{9}{2})^{-}$ states in $^{59}$Fe has been measured via the $^{59}$Co($n,p$) reaction at 198 MeV \cite{ALF93}. The excitation-energy resolution was 0.9 MeV and the data were presented in 1-MeV wide bins. GT strength recorded below $E_{x}=0$ MeV were added to the 0-1 MeV bin for the present study. The experimental strength distribution is shown in Fig. \ref{bgtco59}(a).
Figs. \ref{bgtco59}(b-d) show the theoretical calculations for the GT strength distribution. Fig. \ref{bgtco59}(e) shows the running sums of $B(GT)$ as a function of excitation energy.

A reasonable correspondence between the data and the theoretical calculations using the shell models is found. The calculation with the KB3G interaction has a relatively strong transition to a state at 2.4 MeV, which is weaker in the calculations with the GXPF1a interaction. In both shell-model calculations the summed strength up to an excitation energy of 6 MeV is about 50\% higher than observed in the data. The discrepancy in the summed strength becomes smaller at higher excitation energies, because of the relatively large amount of GT strength found above 6 MeV in the data. In the QRPA calculation a strong transition to a state at 1.5 MeV is predicted, but it is not present in the data. The remainder of the strength appears at around 6 MeV, about 2 MeV higher than where the bulk of the GT strength was found in the experiment. The summed strength up to 10 MeV is overestimated by about 60\% in the QRPA calculation.

\begin{figure}
\includegraphics[scale=0.8]{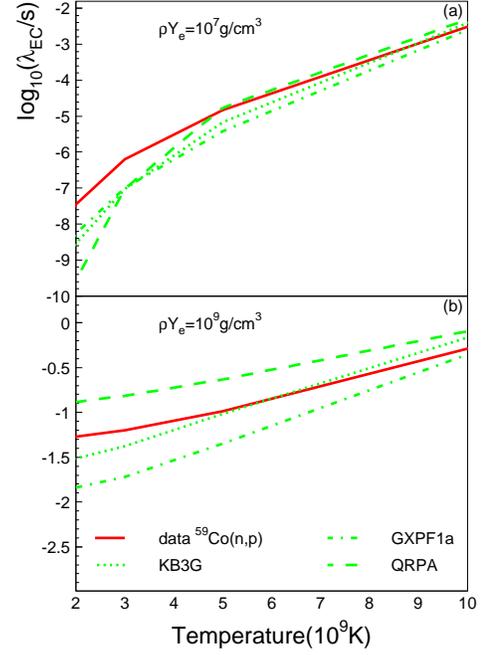}
\caption{Electron-capture rates for $^{59}$Co($e^-,\nu_e$)$^{59}$Fe based on experimental and theoretical GT strengths plotted in Fig. \ref{bgtco59},
as a function of temperature at two densities, $\rho$Y$_e$=$10^7$ g/cm$^3$ (a)
and $\rho$Y$_e$=$10^9$ g/cm$^3$ (b).  \label{ecco59}}
\end{figure}

At $\rho$Y$_e$=$10^7$ g/cm$^3$, and at temperatures below $5\times10^{9}$ K, the EC rates calculated based on the experimentally extracted GT strength distribution are higher than those deduced from the theoretical strength distributions (see Fig. \ref{ecco59}(a)). This could, in part, be due to the placement of some strength at artificially low excitation energies in the experimental distribution as a consequence of the limited resolution. Even the strong transition to the state at 1.5 MeV in the QRPA calculations does not play a large role at these low densities and temperatures. However, at $\rho$Y$_e$=$10^9$ g/cm$^3$ (see Fig. \ref{ecco59}(b)), this transition dominates the EC rate, which then exceeds those based on the data and the shell-model calculations. With slightly more strength placed at lower excitation energies, the EC rates based on the shell-model calculation with the KB3G interaction provides a better match to the EC rates deduced from the experimental strength distribution than the EC rates based on the shell-model calculation with the GXPF1a interaction, except at the highest stellar temperatures.

\subsection{$^{60}$Ni$\rightarrow^{60}$Co \label{SECbgtni60}}

\begin{figure}
\includegraphics[scale=0.95]{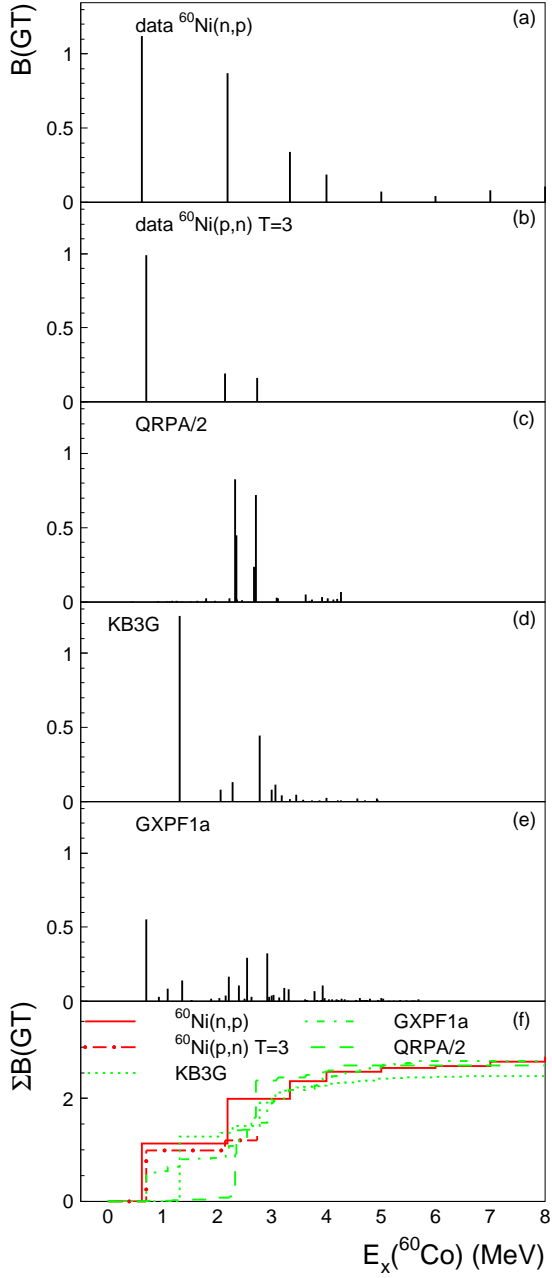}
\caption{\label{bgtni60}GT transitions from $^{60}$Ni to $^{60}$Co: (a) from ($n$,$p$) data, (b) from $T=3$ states observed in ($p$,$n$) data, (c) from calculations in QRPA (divided by a factor of 2), (d) from shell-model calculations using the KB3G interaction and (e) from shell-model calculations using the GXPF1a interaction. In (f), running sums of $B(GT)$ as a function of excitation energy are plotted.}
\end{figure}

The GT strength distribution for transitions from the $^{60}$Ni$(0^{+})$ ground state to $1^{+}$ states in $^{60}$Co has been measured via the $^{60}$Ni($n,p$) reaction at 198 MeV \cite{WIL95}. The energy resolution was 860 keV and the strength distribution was presented in 1-MeV wide excitation-energy bins. The spectrum below $E_{x}(^{60}$Co$)=3.5$ MeV was reanalyzed in Ref. \cite{ANA08} and in the present work the 3-Gaussian fit presented in Fig. 6(b) of Ref. \cite{ANA08} was used to represent the data. For  $E_{x}(^{60}$Co$)>3.5$ MeV, the extracted strengths from Ref. \cite{WIL95} were used. The strength distribution is shown in Fig. \ref{bgtni60}(a). Further experimental information is available from a $^{60}$Ni($p,n$) experiment at 134 MeV \cite{ANA08}, in which GT transitions to states in $^{60}$Cu at high excitation energies were studied. These states are the $T=3$ analogs of the low-lying $1^{+}$ states in $^{60}$Co. Here, we used the extracted $B(GT)$ values from a 3-Gaussian fit (see Fig. 6(a) of Ref. \cite{ANA08}) as shown in Fig. \ref{bgtni60}(b). Since the $T=3$, $1^{+}$ states in $^{60}$Cu are situated in a continuum of $T=1$ and $T=2$ states, extraction of higher-lying GT strengths from the $^{60}$Ni($p,n$) experiment was not possible.

Figs. \ref{bgtni60}(c,d,e) show the theoretical calculations for the GT strength distribution from the QRPA (divided by a factor of 2) and shell-model calculations in truncated $pf$ model space (5 holes in the $f_{7/2}$ shell) using the KB3G and GXPF1a interactions, respectively. Fig. \ref{bgtni60}(f) shows the running sums of $B(GT)$ as a function of excitation energy.

Discrepancies observed between the two available data sets for the transitions to states below 3 MeV make it difficult to draw strong conclusions on the quality of the theoretical calculations. Nevertheless, several observation can be made. In the shell-model calculations with the KB3G interaction, a strong transition to a state at an excitation energy of 1.3 MeV is predicted, with a strength comparable to that of the transition to the state observed at 0.7 MeV in both data sets. In the calculation with the GXPF1a interaction, the excitation energy of this state is very close to the experimental value, but the strength lower by about a factor of 2. Several weaker transitions appear to states at excitation energies just above the strong transition in the calculation with the GXPF1a interaction. The summed GT strengths up to an excitation energy of 8 MeV in the two shell-model calculations are consistent with that found in the $^{60}$Ni($n$,$p$) data. In the QRPA calculations almost all strength is associated with transitions to states with excitation energies between 2 and 3 MeV and the summed strength exceeds that extracted from the data by about a factor of two.

\begin{figure}
\includegraphics[scale=0.8]{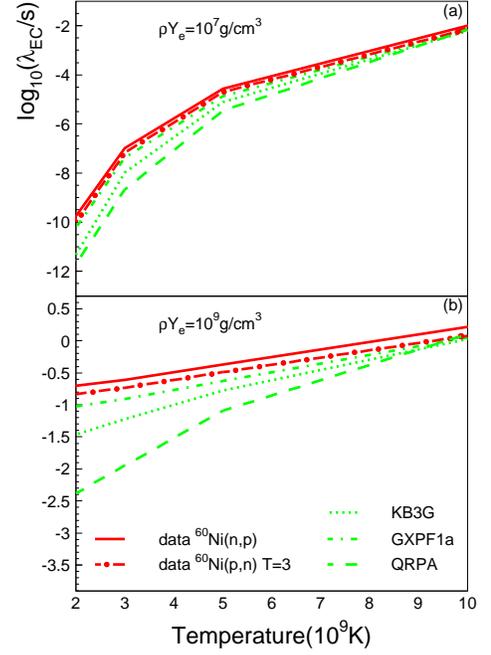}
\caption{Electron-capture rates for $^{60}$Ni($e^-,\nu_e$)$^{60}$Co for experimental and theoretical GT strengths plotted in Fig. \ref{bgtni60},
as a function of temperature at two densities, $\rho$Y$_e$=$10^7$ g/cm$^3$ (a)
and $\rho$Y$_e$=$10^9$ g/cm$^3$ (b).  \label{ecni60}}
\end{figure}

Because $Q_{EC}=-3.334$ MeV. the EC rates on $^{60}$Ni are dominated by the GT transitions to the final states at low excitation energies. Therefore, in spite of the fact that no strength could be extracted from the $^{60}$Ni($p,n$) data at high excitation energies, the difference between the EC rates calculated based on the $^{60}$Ni($n,p$) and $^{60}$Ni($p,n$) data sets are small, as shown in Fig. \ref{ecni60}. Since the location of the final state at the lowest excitation energy is best reproduced in the shell-model calculation with the GXPF1a interaction, the deduced EC rates are also close to those deduced from the data. Although partially mitigated by the overestimated strength of the transition to that state, the EC rates based on the GT strength distribution calculated in the shell-model with the KB3G interaction is lower because the state is placed at higher excitation energies. For the rates based on the QRPA calculations, that effect is enhanced since most of the strength is located at even higher excitation energies. At higher densities and temperatures, where the rates become more sensitive to the total GT strength present in the spectra, the experimental and theoretical rate estimates converge.

\subsection{$^{62}$Ni$\rightarrow^{62}$Co \label{SECbgtni62}}

\begin{figure}
\includegraphics[scale=0.95]{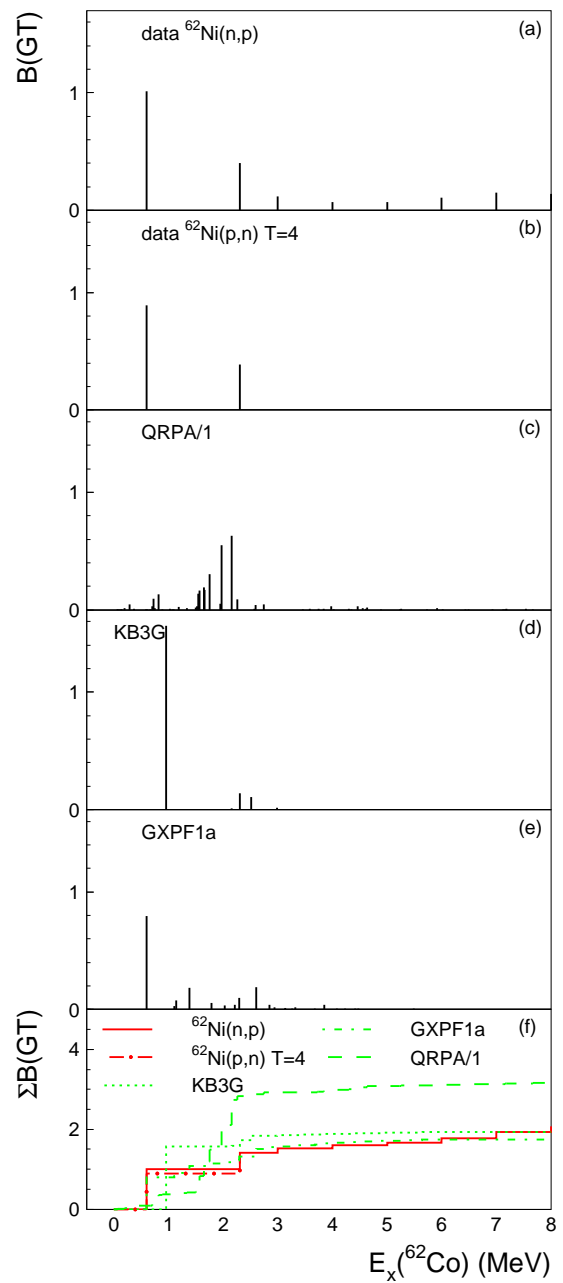}
\caption{\label{bgtni62}GT transitions from $^{62}$Ni to $^{62}$Co: (a) from ($n$,$p$) data, (b) from $T=4$ states observed in ($p$,$n$) data, (c) from calculations in QRPA (divided by a factor of 2), (d) from shell-model calculations using the KB3G interaction and (e) from shell-model calculations using the GXPF1a interaction. In (f), running sums of $B(GT)$ as a function of excitation energy are plotted.}
\end{figure}

Experimental data available for GT transitions from the $^{62}$Ni($0^{+}$) ground state to $1^{+}$ states in $^{62}$Co are similar in nature to those available for $^{60}$Ni. $^{62}$Ni($n,p$) data at 198 MeV \cite{WIL95} provides the GT strength distribution extracted via a multipole decomposition analysis in bins of 1 MeV. Ref. \cite{ANA08} gives a reanalysis of the low-lying part ($E_{x}(^{62}$Co$)<2.5$ MeV) of the $^{62}$Ni($n,p$) excitation-energy spectrum (see Fig. 5(d) of Ref. \cite{ANA08})  and the deduced $B(GT)$ values were used to generate the GT strength distribution shown in Fig. \ref{bgtni62}(a). For $E_{x}(^{62}$Co$)>2.5$, the GT strengths extracted from the multipole decomposition analysis in Ref. \cite{WIL95} were used.
Further experimental information is available from a $^{62}$Ni($p,n$) experiment at 134 MeV \cite{ANA08}. GT transitions to states in $^{62}$Cu at high excitation energies were studied. These states are the $T=4$ analogs of the low-lying $1^{+}$ states in $^{62}$Co. Here, we used the extracted $B(GT)$ values from a 2-Gaussian fit (see Table II of Ref. \cite{ANA08}) as shown in Fig. \ref{bgtni62}(b). Since the $T=4$, $1^{+}$ states in $^{62}$Cu are situated in a continuum of $T=2$ and $T=3$ states, extraction of higher-lying GT strengths from the $^{62}$Ni($p,n$) experiment was not possible. Figs. \ref{bgtni62}(c-e) show the theoretical calculations for the GT strength distribution. Fig. \ref{bgtni62}(f) shows the running sums of $B(GT)$ as a function of excitation energy.

The experimental results from the two sources of data are consistent in the excitation-energy range where both are available. Relatively little GT strength was reported at excitation energies above 3 MeV in the analysis of the $^{62}$Ni($n,p$) data. Both shell-model calculations correctly predict a strong transition to a final state at low excitation energies. The predicted excitation energy in the calculation with the KB3G interaction is too high by about 0.4 MeV and 50\% too strong compared to the experimental strength. Although a little too weak (by about 20\%), the calculation with the GXPF1a interaction predicts the excitation energy correctly. Both shell-model calculations predict little strength at higher excitation energies and in the case of the KB3G calculation some of the excess strengths predicted for the transition to the first $1^{+}$ state is balanced by the reduced amount of strength predicted at higher excitation energies. Consequently, in both calculations the total GT strength is in close correspondence with the data. In the QRPA calculation, a strong state at low excitation energies is absent and the bulk of the strength, in excess of the data by about a factor of 2, is concentrated in a narrow excitation energy window centered around 2 MeV. There are, however, about 10 very weak transitions to states below 1.0 MeV (with excitation energies of as low as 0.06 MeV), which are not all visible in Fig. \ref{bgtni62}(c).

Since $Q_{EC}=-5.862$ MeV, the EC rates are strongly dominated by transitions to final states at low excitation energies, even at the higher density of $10^{9}$ g/cm$^{3}$. For the strength distributions extracted from the data and the strength distributions calculated in the shell-models, the response is dominated by the strong transition to the first $1^{+}$ state.
The slightly higher excitation energy of this state in the calculation with the KB3G interaction reduces the deduced EC rate compared to that deduced from the calculation with the GXPF1a interaction. However, this reduction is countered by the slightly higher strength for the transition to this state in the calculation with the KB3G interaction.

\begin{figure}
\includegraphics[scale=0.8]{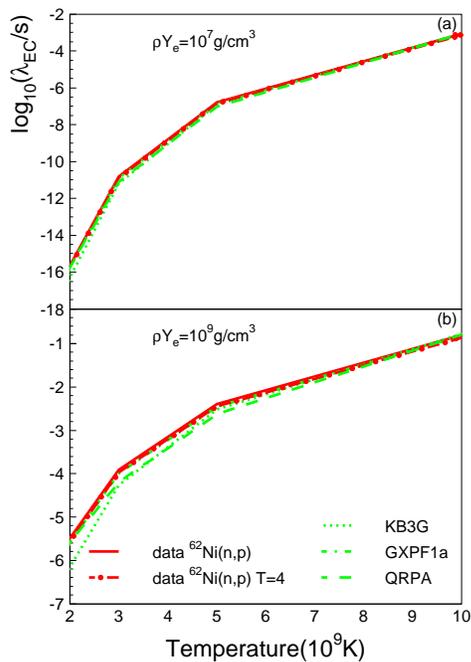}
\caption{Electron-capture rates for $^{62}$Ni($e^-,\nu_e$)$^{62}$Co for experimental and theoretical GT strengths, plotted
as a function of temperature at two densities, $\rho$Y$_e$=$10^7$ g/cm$^3$ (a)
and $\rho$Y$_e$=$10^9$ g/cm$^3$ (b).  \label{ecni62}}
\end{figure}

The opposite effect is seen for the EC rate based on the QRPA calculation due to several very weak transitions to states below 1 MeV. The effects of the small transition strengths on the EC rates are countered by the low-excitation energies of the final states and the deduced EC rates are relatively close to the other EC rates, as shown in Fig. \ref{ecni62}.

\subsection{$^{64}$Ni$\rightarrow^{64}$Co \label{SECbgtni64}}

\begin{figure}
\includegraphics[scale=0.95]{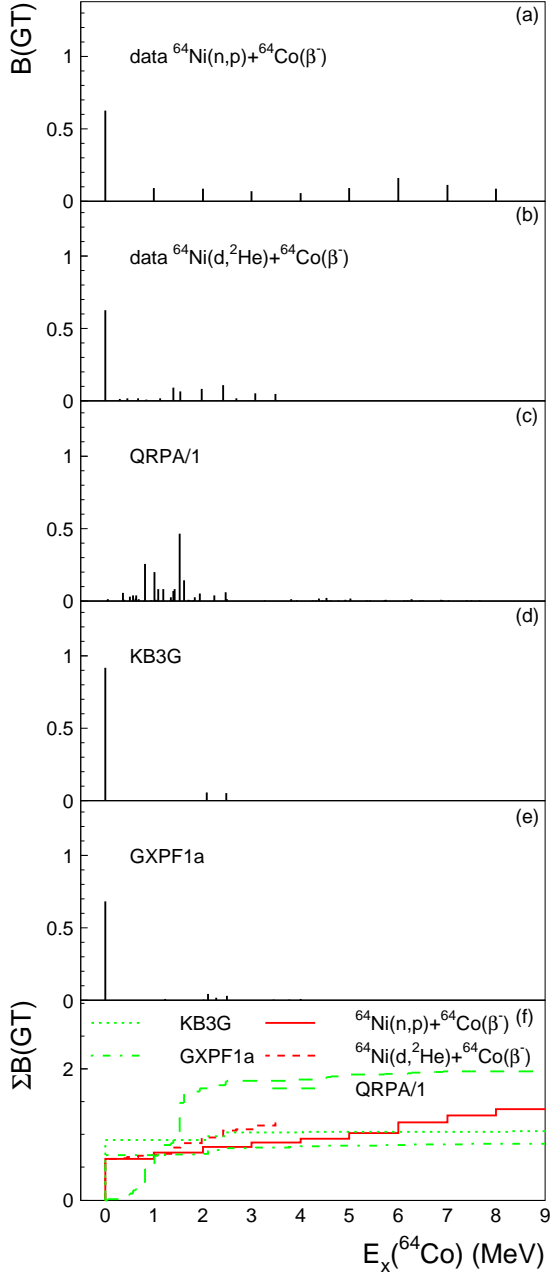}
\caption{\label{bgtni64}GT transitions from $^{64}$Ni to $^{64}$Co: (a) from ($n$,$p$) and $\beta$-decay data, (b) from ($d$,$^{2}$He) and $\beta$-decay data, (c) from calculations in QRPA, (d) from shell-model calculations using the KB3G interaction and (e) from shell-model calculations using the GXPF1a interaction. In (f), running sums of $B(GT)$ as a function of excitation energy are plotted.}
\end{figure}

Data on GT transitions from $^{64}$Ni are available from $^{64}$Ni($n,p$) \cite{WIL95} and $^{64}$Ni($d$,$^{2}$He) \cite{POP07} experiments. In addition, the $B(GT)$ for the transition from the $^{64}$Ni($0^{+}$) ground state to the $^{64}$Co($1^{+}$) ground state can be deduced from a $^{64}$Co($\beta^{-}$) experiment ($B(GT)=0.627$) \cite{RAH74}. The resolution in the $^{64}$Ni($n,p$) experiment was 800 keV and the GT strength distribution was extracted through a multipole decomposition analysis in 1-MeV wide excitation-energy bins. In the present study, these data were combined with the GT strength extracted for the $^{64}$Co($\beta^{-}$) transition to the ground state, as shown in Fig. \ref{bgtni64}(a).
The energy resolution achieved in the $^{64}$Ni($d$,$^{2}$He) experiment was 110 keV, allowing for extraction of the $B(GT)$ values on a state-by-state basis up to an excitation energy in $^{64}$Co of 1.5 MeV and in narrow energy-bins up to 3.7 MeV (Fig. \ref{bgtni64}(b)). The $^{64}$Co($\beta^{-}$) data was used in Ref. \cite{POP07} to calibrate the extraction of GT strengths. The $B(GT)$ for the transition from the $^{64}$Ni($0^{+}$) ground state to the $^{64}$Co($1^{+}$) ground state is therefore by definition equal to that deduced from the $\beta^{-}$ experiment \cite{RAH74}.

Figs. \ref{bgtni64}(c-e) show the theoretical calculations for the GT strength distribution. Fig. \ref{bgtni64}(f) shows the running sums of $B(GT)$ as a function of excitation energy.

The dominant GT transition to the ground state is well reproduced (difference of less than 10\%) by the shell-model calculation using the GXPF1a interaction and overestimated (by about 40\%) in the calculation using the KB3G interaction. Both shell-model calculations predict less strength at higher excitation energies than observed in the data. The results from the high-resolution ($d$,$^{2}$He) experiment suggest that the extraction of GT strengths from the ($n$,$p$) data using a MDA led to the loss of a small fraction of GT strength, at least up to 3.7 MeV. Since GT strengths were not extracted from the ($d$,$^{2}$He) data above that excitation energy, is not clear whether that trend would continue up to the higher excitation energies. Even under the assumption that no strength was lost in the analysis of the $(n,p)$ data, the total GT strength predicted in the shell-model calculations is about 30\% lower than observed in the data. The QRPA calculations fail to reproduce the strong transition to the ground state of $^{64}$Co; most of the GT strength is located at excitation energies between 1 and 2 MeV and the total strength is in excess of the experimental strength by about a factor of two.

\begin{figure}
\includegraphics[scale=0.8]{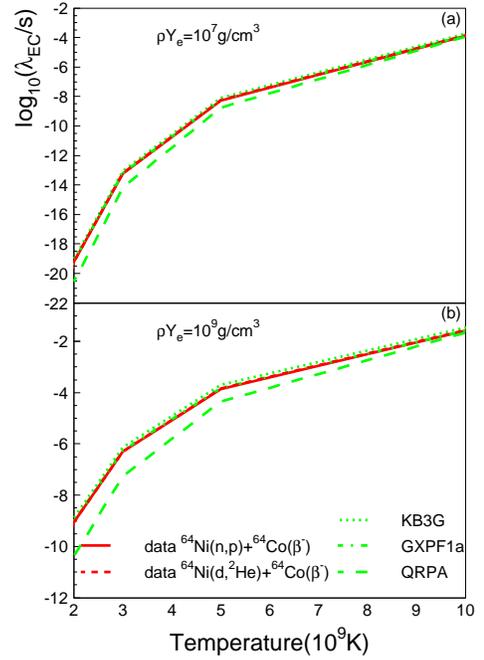}
\caption{Electron-capture rates for $^{64}$Ni($e^-,\nu_e$)$^{64}$Co for experimental and theoretical GT strengths plotted in Fig. \ref{bgtni64},
as a function of temperature at two densities, $\rho$Y$_e$=$10^7$ g/cm$^3$ (a)
and $\rho$Y$_e$=$10^9$ g/cm$^3$ (b).
\label{ecni64}}
\end{figure}

Since $Q_{EC}=-7.818$ MeV, the EC rates on $^{64}$Ni are very small and dominated by the transition to the ground state. Since the shell-model calculation with the GXPF1a interaction provides a good match to the data, the EC rates deduced from this model are very close to those based on the experimental GT strength distribution (see Fig. \ref{ecni64}). The EC rates based on the shell-model calculation with the KB3G interaction is too high by about 40\% due to the overestimation of the GT strength of the transition to the ground state of $^{64}$Co; that discrepancy is not visible due to the logarithmic scales of Fig. \ref{ecni64}. EC rates based on the GT strength distribution calculated in QRPA are too low, in particular at low temperatures, due to the absence of the strong transition to the ground state.

\subsection{$^{64}$Zn$\rightarrow^{64}$Cu \label{SECbgtzn64}}

\begin{figure}
\includegraphics[scale=0.95]{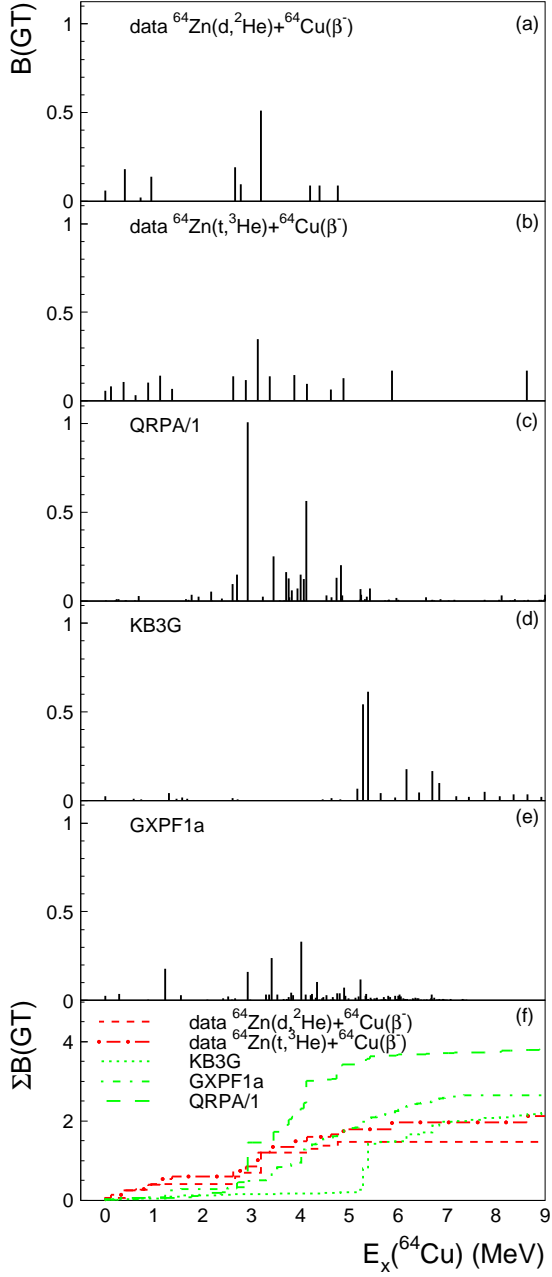}
\caption{\label{bgtzn64}GT transitions from $^{64}$Zn to $^{64}$Cu: (a) from ($d$,$^{2}$He) and $\beta$-decay data, (b) from ($t$,$^{3}$He) and $\beta$-decay data, (c) from calculations in QRPA, (d) from shell-model calculations using the KB3G interaction and (e) from shell-model calculations using the GXPF1a interaction. In (f), running sums of $B(GT)$ as a function of excitation energy are plotted.}
\end{figure}

Data on GT transitions from the $^{64}$Zn($0^{+}$) ground state to  $^{64}$Cu($1^{+}$) states are available from $^{64}$Zn($d$,$^{2}$He) \cite{GRE08} and $^{64}$Zn($t$,$^{3}$He) \cite{HIT09} experiments. Additionally, the $B(GT)$ for the transition from the $^{64}$Zn($0^{+}$) ground state to the $^{64}$Cu($1^{+}$) ground state can be deduced from $^{64}$Cu($\beta^{-}$) experiments ($B(GT)=0.059$) \cite{SIN07}.
The energy resolution achieved in the $^{64}$Zn($d$,$^{2}$He) experiment was 115 keV, allowing for extraction of the $B(GT)$s on a state-by-state basis up to an excitation energy in $^{64}$Cu of 5 MeV, as shown in Fig. \ref{bgtzn64}(a). The $^{64}$Cu($\beta^{-}$) data was used in Ref. \cite{GRE08} to calibrate the extraction of GT strengths and the $B(GT)$ for the transition from the $^{64}$Zn($0^{+}$) ground state to the $^{64}$Cu($1^{+}$) ground state was, therefore, by definition equal to that deduced from the $\beta^{-}$ experiment.
The energy resolution achieved in the $^{64}$Zn($t$,$^{3}$He) experiment was 280 keV and a multipole decomposition analysis was performed for 250-keV wide excitation-energy bins up to 9 MeV. The extraction of the GT strengths in the analysis did not rely on a calibration of the unit cross section using the $^{64}$Cu($\beta^{-}$) data. However, the $B(GT)$ for the ground-state to ground-state transition was set equal to that obtained from the $^{64}$Cu($\beta^{-}$) data, as shown in Fig. \ref{bgtzn64}(b).
Figs. \ref{bgtzn64}(c-e) show the theoretical calculations for the GT strength distribution.  Fig. \ref{bgtzn64}(f) shows the running sums of $B(GT)$ as a function of excitation energy.

Compared to the other nuclei studied in this work, the GT strength distributions calculated in the shell-models provide a poor description of the data. The calculation with the KB3G interaction places the bulk of the strength at much too high excitation energies and the calculation with the GXPF1a interaction also places too little strength at low excitation energies. The strength calculation based on the QRPA calculations suffers from a similar problem. In addition, it overestimates the total strength by nearly a factor two. Both shell-model calculations underestimate the GT strength for the transition to the ground state; by 50\% for the GXPF1a and and by 80\%  for the KB3G interaction. The GT strength for the transition to the ground state is underestimated by a factor of 16 in the QRPA calculations and not visible in the plotted spectrum.

Not surprisingly, the EC rates based on the theoretical calculations underestimate those based on the experimental strength distributions. Since $Q_{EC}=-1.09$ MeV, GT strengths at relatively high excitations energies can contribute significantly, especially at the higher density. This somewhat reduces the discrepancies between the EC rates based on the data and the theory.

\begin{figure}
\includegraphics[scale=0.8]{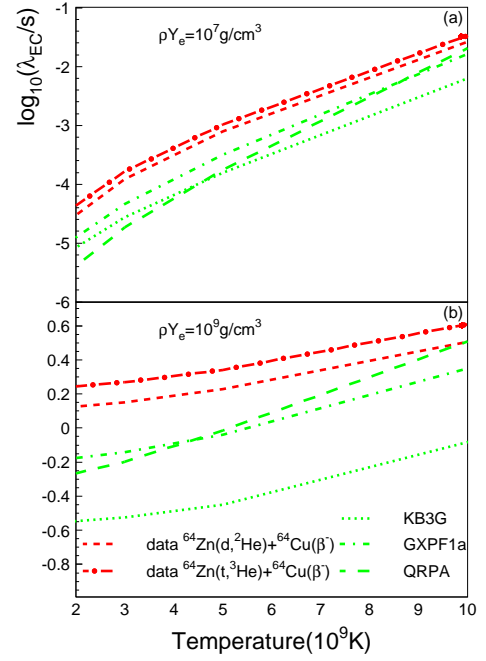}
\caption{Electron-capture rates for $^{64}$Zn($e^-,\nu_e$)$^{64}$Cu for experimental and theoretical GT strengths shown in Fig. \ref{bgtzn64}, plotted
as a function of temperature at two densities, $\rho$Y$_e$=$10^7$ g/cm$^3$ (a)
and $\rho$Y$_e$=$10^9$ g/cm$^3$ (b).  \label{eczn64}}
\end{figure}

\section{Discussion \label{SECdiscussion}}

Based on the discussion of the GT strength distribution for individual nuclei in the previous section, one can conclude that the shell-model calculations with the KB3G and GXPF1a interaction describe the main features of experimental GT strength distributions. The exceptions are the cases of $^{64}$Zn and $^{45}$Sc. For the latter case, high-resolution, background-free data are needed to better assess the discrepancies between experiment and shell-model theory.
For nuclei with mass number $A\geq 56$, the GT strength distributions calculated with the GXPF1a interaction exhibit more fragmentation and are somewhat broader than those calculated with the KB3G interaction. With the latter interaction, more strength is placed at low-excitation energies, and less at higher excitation energies. Similar differences between these two shell-model calculations also became apparent in the recent measurement of the GT strength distribution from unstable $^{56}$Ni \cite{SAS11}. In general, the calculations performed in the QRPA formalism do not describe the details of the strength distributions well, and the total strength is overestimated by up to a factor of two.

\begin{figure*}
\centering
\includegraphics[scale=1.0]{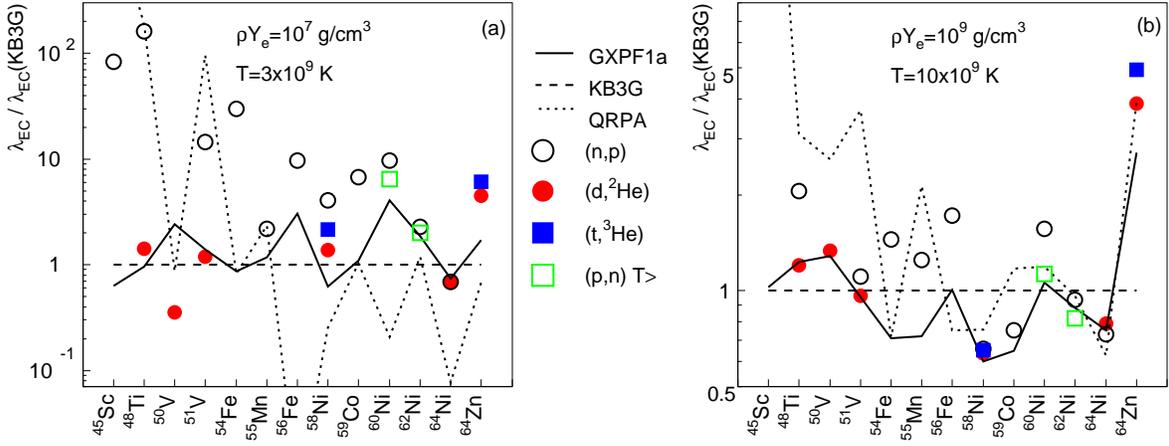}
\caption{(color online) Comparison of EC rates for 13 $pf$-shell nuclei calculated on the basis of theoretical (connected by lines) and experimental (indicated by markers) GT strength distributions. All rates are plotted relative to those calculated by in the shell-model with the KB3G interaction (for which the relative rate thus equals unity).
The relative EC rates are shown for two combinations of stellar density and temperature: (a) $\rho$Y$_e$=$10^7$ g/cm$^3$ and $T=3{\times}10^9$ K and (b)  $\rho$Y$_e$=$10^9$ g/cm$^3$ $T=10{\times}10^9$ K.\label{rates}}
\end{figure*}

To gain more quantitative insight in the impact of the differences between the experimental and theoretical GT strength distributions on the derived EC rates, the EC rates on the ground state for each of the 13 $pf$-shell nuclei investigated in this work are compared in Fig. \ref{rates} for two specific stellar density-temperature combinations (cases I and II: see section \ref{SECbgt}).  In Fig. \ref{rates}(a) rates at a density of $\rho$Y$_e$=$10^7$ g/cm$^3$ and a temperature of $T=3{\times}10^9$ K are compared (case I), and in Fig. \ref{rates}(b), rates at a density of  $\rho$Y$_e$=$10^9$ g/cm$^3$ and a temperature of $T=10{\times}10^9$ K are compared (case II).
Depending on the value of $Q_{R}=E_{1}-Q_{EC}$, the magnitudes of the absolute EC rates varies strongly between the cases studied here. Therefore, we chose to plot relative EC rates, and used the rates derived from the theoretical strength distributions using the shell-model with the KB3G interaction as a reference.

For case I (Fig. \ref{rates}(a)), the EC rates are very sensitive to the details of the GT strength distribution at low excitation energies. Of the nuclei studied, only in the cases of $^{45}$Sc and $^{64}$Zn is $\varepsilon_{F}(T=0)=1.2$ MeV$>Q_{R}$ for the transition to the states at the lowest excitation energies. In all other cases EC can only proceed at finite temperature. The placement of the states thus becomes critically important because the phase space for EC decreases rapidly with increasing excitation energy.

Since the resolution of the GT strength distributions extracted from the $(n,p)$ experiments is limited, strength might be placed at artificially low excitation energies, resulting in an overestimate of the EC rates. This is clearly visible in Fig. \ref{rates}(a) for the cases of $^{48}$Ti, $^{51}$V and $^{58}$Ni, for which $(n,p)$ and high-resolution data are available.  For the EC rates derived from GT strengths extracted from the $^{60,62,64}$Ni($n$,$p$) data, such artificial inflation does not occur as the GT strength at low excitation energies is completely dominated by a transition to a single state. Even with poor experimental resolution, the excitation energy of this state could be established with little uncertainty.

Ignoring the cases for which the excitation energies of the states at low excitation energies could not be well determined, it is clear that the deviations between the EC rates based on the shell-model calculations and the EC rates based on experimental GT strength distributions are relatively small compared to the deviations observed for the rates based on the QRPA calculations.
To quantify the differences, we defined an average deviation ($\overline{\Delta_{EC}}$) between the EC rates based on the charge-exchange data ($\lambda^{exp}$) and the theory ($\lambda^{th}$), as a fraction of the experiment values as:
\begin{equation}
\label{eq:rd}
\overline{\Delta_{EC}}=\frac{1}{N} \sum_{i=1}^{N} \frac{\lambda_{i}^{th}-\lambda_{i}^{exp}}{\lambda_{i}^{exp}},
\end{equation}
where the sum refers to all nuclei for which high resolution data are available, or the GT strength distribution at low excitation energy is otherwise well-defined: $^{48}$Ti, $^{51}$V, $^{58}$Ni, $^{60}$Ni, $^{62}$Ni, $^{64}$Ni and $^{64}$Zn were used. Although high-resolution data are available for $^{50}$V, we excluded them from this analysis because of the ambiguity related to the strength of the GT transition to the first $6^{+}$ state in the experimental analysis.

Since $\overline{\Delta_{EC}}$ can be small when averaged over several nuclei if positive and negative deviations cancel out, we also defined $\overline{|\Delta_{EC}}|$:
\begin{equation}
\label{eq:ad}
\overline{|\Delta_{EC}|}=\frac{1}{N} \sum_{i=1}^{N} \frac{|\lambda_{i}^{th}-\lambda_{i}^{exp}|}{\lambda_{i}^{exp}},
\end{equation}
which represents the average value of the absolute deviation between the EC rates based on the charge-exchange data (using the same 7 nuclei as for the calculation of $\overline{\Delta_{EC}}$) and the theory, as a fraction of the EC rate based on the data.
The values of $\overline{\Delta_{EC}}$ and $\overline{|\Delta_{EC}|}$ are presented in Table \ref{tab:rates} for each of the three theoretical models tested in this work, and for both temperature-density combinations defined above.

\begin{table*}
  \caption{\label{tab:rates}Average deviations between EC rates calculated based on theoretical GT strength distributions and EC rates based on GT strengths extracted from charge-exchange experiments, relative to the experimental values, for two stellar density-temperature combinations. Only EC rates on the ground states of 7 nuclei ($^{48}$Ti, $^{51}$V, $^{58}$Ni, $^{60}$Ni, $^{62}$Ni, $^{64}$Ni and $^{64}$Zn) in the $pf$ shell for which high-resolution data were available are considered, or the strength distribution at low excitation energies in the daughter nucleus were otherwise well-defined. The left-hand side of the table refers to deviations at $\rho$Y$_e$=$10^7$ g/cm$^3$ and a temperature of $T=3{\times}10^9$ K (case I) and the right-hand-side of the table refers to deviations at $\rho$Y$_e$=$10^9$ g/cm$^3$ and $T=10{\times}10^9$ K (case II).}
  \begin{ruledtabular}
  \begin{center}
    \begin{tabular}{ccccccc}
    & \multicolumn{3}{c}{I: $\rho$Y$_e$=$10^7$ g/cm$^3$ $T=3{\times}10^9$ K  } & \multicolumn{3}{c}{II: $\rho$Y$_e$=$10^9$ g/cm$^3$ $T=10{\times}10^9$ K} \\
    \cline{2-4}\cline{5-7}
    \noalign{\smallskip}
          & GXPF1a & KB3G  & QRPA  & GXPF1a & KB3G  & QRPA \\
    \cline{2-2}\cline{3-3}\cline{4-4}\cline{5-5}\cline{6-6}\cline{7-7}
    \noalign{\smallskip}
    $\overline{\Delta_{EC}}$ & -0.24  & -0.34 & 29.    & -0.05 & 0.01 & 0.66 \\ \\
    $\overline{|\Delta_{EC}|}$ & 0.31  & 0.47  & 30.    & 0.08  & 0.30  & 0.72 \\
    \end{tabular}
  \end{center}
  \end{ruledtabular}
\end{table*}

For case I, the rates based on the shell-model calculations with the GXPF1a and KB3G interactions underestimate the rates based on the data on average by approximately $30$\% ($\overline{\Delta_{EC}}\approx 0.3$). The average absolute deviation ($\overline{|\Delta_{EC}|}$) is slightly lower for the calculation using the GXPF1a interaction compared to the calculation using the KB3G interaction. However, given the uncertainties in the experimental strengths and rates, one cannot draw very strong conclusions based on this difference. The rates based on GT strength distributions produced in QRPA deviate strongly from those based on the data and it is clear that at low stellar densities and temperatures, the QRPA rates should be used with caution when modeling late stellar evolution.

For case II, derived EC rates are less sensitive to the details of the strength distribution and the differences between the rates calculated based on theoretical and experimental strength distributions become smaller, as can be seen in Fig. \ref{rates}(b) (notice the difference in the scale of the axis depicting the rates from Fig. \ref{rates}(a)).
The EC rates derived from $(n,p)$ data, for which the GT strength at low-excitation energies could not be accurately placed, are still too high compared to those derived from the high resolution data. However, the differences have reduced to less than a factor of two.
The EC rates derived from the shell-model calculations both match well with those derived from the data. The average deviation ($\overline{\Delta_{EC}}$) between the EC rates derived from these calculations and the data for the 7 nuclei, for which the strength at low excitation energies could be well-determined from the experiments, is less than 10\% (see Table \ref{tab:rates}). For the calculations with the GXPF1a interaction, the average absolute deviation ($\overline{|\Delta_{EC}|}$) is also below 10\%. For the strength extracted from the calculations with the KB3G interaction the average absolute deviation is 30\%.

Since the EC rates at higher densities and temperatures are more sensitive to the overall GT strength, rather than the exact location of the daughter states at low excitation energies, the EC rates calculated based on the GT strength distributions from the QRPA formalism deviate much less from the rates based on the charge-exchange data than at the lower density and temperature. On average, the rates are about 70\% too high, which roughly corresponds with the average overestimation of the total GT strength in this theory, in comparison to the data.

\section{Summary and Conclusion\label{conclusion}}
We have evaluated 25 sets of data available from ($n$,$p$), ($d$,$^{2}$He), ($t$,$^{3}$He) and ($p$,$n$) experiments, providing information about GT strength distributions in the $\Delta T_{z}=+1$ direction for 13 nuclei in the $pf$ shell. Three sets of theoretical strength distributions (two of which were calculated in shell-models using the KB3G and GXPF1a interactions and one was based on QRPA calculations using ground-state deformation parameters and masses from the finite-range droplet model) were tested against the experimental results. In addition, EC rates of relevance to late stellar evolution were derived from the experimental and theoretical GT strength distributions and compared as well.

The GT strength distributions calculated in the shell-models do about equally well in reproducing the data, whereas large differences were observed for the QRPA calculations. Consequently, derived EC rates from the shell-model calculations are also much closer to the EC rates derived from the experimental GT strength distributions than those calculated on basis of the QRPA framework. In order to perform quantitative comparisons between experimental and theoretical EC rates, we only used nuclei for which high-resolution data exist or for which the location of daughter states at low excitation energies could be established with small uncertainty. At relatively low stellar densities and temperatures ($\rho$Y$_e$=$10^7$ g/cm$^3$ and $T=3{\times}10^9$ K), the EC rates derived from GT strengths calculated in the shell models were on average about 30\% higher than the EC rates derived from the experimental GT strength distributions. Average absolute deviations were less than 50\%.
At higher densities and temperatures ($\rho$Y$_e$=$10^9$ g/cm$^3$ and $T=10{\times}10^9K$), these deviations became smaller: the average absolute deviation between the EC rates based on the shell-model calculations using the GXPF1a interaction and the data was less than what could be attributed to experimental uncertainties (10\%). For the calculations with the KB3G interaction that deviation was slightly higher (30\%), but when the sign of the deviations were taken into account, the average deviation reduced to below the uncertainties. EC rates calculated on the basis of GT strength distributions calculated in QRPA exhibited much larger deviations, in particular at low stellar densities. Usage of EC rates based on these theoretical calculations is better avoided in simulations of late stellar evolution that are very sensitive to these rates.

The results from this work can serve as a benchmark for future theoretical studies that aim to provide estimates for EC rates of relevance for stellar evolution.
In addition, they give a first estimate for the uncertainties in EC rates that are used in astrophysical simulations which can serve as input for sensitivity studies in stellar-evolution modeling.

Except for the lightest nuclei (near $^{45}$Sc), sufficient data are available to test theoretical predictions for GT strength distributions in stable $pf$-shell nuclei. However, very little is known about the quality of the theoretical predictions and the derived EC rates for $pf$-shell nuclei further away from the valley of stability, and for nuclei beyond the $pf$-shell, since data are scarce and the available theoretical tools more limited. Future experimental programs and advances in theoretical techniques will allow for extension of comparative studies such as the one presented here to these heavier and more exotic nuclei.

\begin{acknowledgments}
This work was supported by the US NSF (PHY-0822648 (JINA), PHY-0606007, PHY-0758099, PHY-1102511), and the Research Corporation for Science Advancement. The authors also wish to thank Gabriel Mart\'{i}nez-Pinedo for pointing out an error in the treatment of the electron-mass in the electron-capture rate calculations.
\end{acknowledgments}

\bibliography{prc}

\end{document}